\documentclass{article}
\usepackage{axodraw}
\usepackage{pstcol}
\usepackage{graphicx}
\newcommand{\bmat}{\left(\begin{array}}
\newcommand{\emat}{\end{array}\right)}
\newcommand{\be}{\begin{equation}}
\newcommand{\ee}{\end{equation}}
\newcommand{\bea}{\begin{eqnarray}}
\newcommand{\eea}{\end{eqnarray}}

\def\NPB#1#2#3{Nucl. Phys. B{#1} (19#2) #3}
\def\PLB#1#2#3{Phys. Lett. B{#1} (19#2) #3}

\def\PRD#1#2#3{Phys. Rev. D{#1} (19#2) #3}

\def\A{{\bf A}}

\def\inte{{\bf Z}}

\def    \be            {\begin{equation}}
\def    \ee            {\end{equation}}
\def    \bea           {\begin{eqnarray}}
\def    \eea           {\end{eqnarray}}
\def\eps{\epsilon}
\def\a{\alpha}
\def\b{\beta}

\def\d{\delta}

\def\k{\kappa}

\def\vt{\vartheta}

\def\la{\lambda}

\newcommand{\matr}[9]{\left(\begin{array}{ccc}{#1}&{#2}&{#3}\\
{#4}&{#5}&{#6}\\{#7}&{#8}&{#9}\end{array}\right)}
\begin{document}
\title{Fermion Masses and Mixing in Four and More Dimensions}

\author{N. Chamoun$^{1}$ and S. Khalil$^{2,3}$\\
{\small $^1$ Physics Department, HIAST, P.O.Box 31983,
Damascus, Syria.}\\
{\small $^2$ Ain Shams University, Faculty of Science, Cairo 11566,
Egypt.}\\
{\small $^3$Department of Mathematics, German University in Cairo-
GUC, new Cairo, Egypt.}}

\maketitle
\begin{abstract}
We give an overview of recent progress in the study of fermion mass
and flavor mixing phenomena. Mass matrix ansatze are considered
within the SM and SUSY GUTs where some predictive frameworks based
on SU(5) and SO(10) are reviewed. We describe a variety of schemes
to construct quark mass matrices in extra dimensions focusing on
four major classes: models with the SM residing on 3-brane, models
with universal extra dimensions, models with split fermions and
models with warped extra dimensions. We outline how realistic
patterns of quark mass matrices could be derived from orbifold
models in heterotic superstring theory. Finally, we address the
fermion mass problem in intersecting D-branes scenarios, and present
models with D6-branes able to give a good quantitatively description
of quark masses and mixing. The role of flavor/CP violation problem
as a probe of new physics is emphasized.
\end{abstract}
\section{Introduction}

The origin of the quark and lepton masses, their mixing and three
generation structure remains the major outstanding problem in
particle physics. It has different aspects, questioning the origin
of the family replication, the fermion mass spectrum, CP violation
in weak interactions, suppression of the flavour changing neutral
currents (FCNC), pattern of neutrino masses and oscillations, etc.
The charged fermion masses and mixing angles derive from Yukawa
couplings, which are arbitrary parameters within the Standard
Model (SM), while the non-vanishing neutrino masses and mixings
are direct evidence for physics beyond the SM. The experimental
values of the fermion masses and mixings provide a best clue to
this new physics. We require explanations for the large mass
ratios between generations: $m_u \ll m_c \ll m_t ; \quad m_d \ll
m_s \ll m_b ; \quad m_e \ll m_{\mu} \ll m_{\tau}$, and for the
large mass splitting within the third (heaviest) generation:
$m_{\tau} \sim m_b \ll m_t$. We need also an explanation for the
smallness of the off-diagonal elements of the quark weak coupling
matrix $V_{CKM}$ and for the tiny neutrino masses and their large
mixings  as recent data suggest.

Grand Unified Theories (GUTs) have the power to fill the gap
between theory and experiment. Indeed,  within this framework the
low energy group proceeds from the spontaneous breakdown of a
single compact group.  The simplest and most attractive grand
unified theories are based on the unitary group $SU(5)$ or the
orthogonal group  $SO(10)$.  Remarkably,  all low energy fermion
quantum numbers find a natural explanation within these theories.
If the grand unified group breaks at very high energies to the
standard model gauge group, an essential requirement is that the
theory should be supersymmetric \cite{cdwref2}. In fact,
supersymmetric theories have been the main extension of the SM for
about 20 years. Not only supersymmetry (SUSY) provides a natural
protection for the weak scale against any large scale as long as
SUSY breaking is around $TeV$ (Hierarchy problem), but also, when
incorporated in GUTs, it can predict the weak mixing angle in
remarkably good agreement with the precise measurements at LEP
\cite{bmsref1} and leads to relations among quark and lepton
masses. The most celebrated one is $m_b=m_\tau$ at the
unification scale which is corroborated by experiment once the
running is taken into account \cite{bmsref3}. In SUSY there are
additional sources of CP violation to the SM source arising from
the complex Yukawa couplings. This is due to the presence of new
CP violating phases which arise from the complexity of the soft
SUSY breaking terms and the SUSY preserving $\mu$-parameter. These
new phases have significant implications and can modify the SM
predictions, whence the experimental data on the CP asymmetries
in K and B systems impose constraints on the SUSY GUT frameworks,
and can help in picking up the right one (see \cite{khalil02} and
references therein).

The other popular scenario for the solution of the hierarchy
problem is large extra dimensions
\cite{ahhswref1,ahhswref2,ahhswref3}, where the observed weakness
of gravity is due to the existence of new spatial dimensions large
compared to the weak scale. The original scenario where the SM
fields reside on a 3-brane, with a low fundamental cut-off and
extra dimensions, allows having flavor physics close to the TeV
scale \cite{ahhsw}. Small Yukawa couplings are generated by
``shining'' badly broken flavor symmetries from distant branes,
and flavor and CP-violating processes are adequately suppressed
by these symmetries. Other scenarios where extra dimensions are
accessible to {\it all\/} the SM fields were proposed and referred
to as universal extra dimensions (UED) \cite{acd}. Here, the
compactification scale can be lower than the previous scenarios
because the Kaluza-Klein (KK) number in the equivalent
four-dimensional theory is conserved, and thus the contributions
to the electroweak observables arise only from loops. One can
study the impact of UED on the values of the CKM parameters and
whether there are interesting phenomenological implications on K
and B decays \cite{kmref7}. Also, we can envisage a scenario where
the SM fields are confined to a thick wall in extra dimensions,
with the fermions localized in specific points in the wall
\cite{kmref21}. In this so called `Split Fermions' scenario,
Yukawa couplings are suppressed due to the exponentially small
overlaps of the fermions wave functions. This provides a
framework for understanding both the fermion mass hierarchy and
proton stability without imposing symmetries, but rather in terms
of higher dimensional geometry. Another mechanism to solve the
hierarchy problem in extra dimensions is the R-S warped
non-factorizable geometry \cite{RS1,RS2}. The weak scale is
generated from a large scale of order of the Planck scale through
an exponential `warp' factor which arises not from gauge
interactions but from the background metric (which is a slice of
AdS$_5$ spacetime). One can then explore the phenomenology
associated with this localized gravity model, and the KK tower of
gravitons have strikingly different properties than in the
factorizable case with large extra dimensions\cite{davoudiasl}.

It is possible to embed the above scenarios within string theory,
where fields confined on a brane are identified with open strings
whose ends are attached to this brane. In fact, the true
resolution to the flavor problem lies in the domain of the
underlying fundamental theory of which the SM would be the low
energy effective theory. Since at present
Superstrings/``M"--theory is the only candidate for a truly
fundamental quantum theory of all interactions, studies of the
flavor structure of the Yukawa couplings within four-dimensional
superstring models are well motivated. In particular, the
couplings of the effective Lagrangian in superstring theory are
in principle calculable and not input parameters, which allows to
address the flavor problem quantitatively. Indeed, the structure
of fermion masses has been studied in a number of semi-realistic
heterotic string models such as orbifolds
\cite{cklref7,cklref8one,cklref8two} which have a beautiful
geometric mechanism to generate a mass hierarchy and the
resulting renormalizable Yukawa couplings can be explicitly
computed  as functions of the geometrical moduli
\cite{cklref9,cklref10one,cklref10two}. With the advent of
Dirichlet D-branes, the phenomenological possibilities of string
theory have widened in several respects, and the flavor problem
within `intersecting D-branes' models \cite{cklref24,cklref27}
seems promising. In these models, chiral fields to be identified
with SM fermions live at different brane intersections and there
is a natural origin for the replication of quark-lepton
generations in that the branes would typically intersect a
multiple number of times giving rise to the family structure.
Moreover, the appearance of hierarchies in Yukawa couplings of
different fermions comes naturally because these couplings are
weighted exponentially with the area of the triangle shape in
whose vertices lie the chiral fields, and thus different triangle
areas corresponding to the various families could generate a
hierarchical structure.

The structure of the review is as follows. We give in section 2 an
overview of the quark-lepton spectrum and the generation of
fermion masses and mixings in the SM. The hierarchy problem and
its supersymmetric solution is outlined in section 3 where flavor
issues are reviewed within the Minimal Supersymmetric Standard
Model (MSSM). Fermion masses in SUSY GUTs are reviewed in section
4 where we examine models based on SU(5) and SO(10). Large extra
dimensions as originally proposed or within the UED picture are
reviewed in section 5 with regard to the flavor hierarchies. The
implications of split fermions and warped geometry on the fermion
masses and mixings are presented in sections 6 and 7
respectively. In section 8 we outline orbifold models studying
the Yukawa structure. Section 9 is devoted to studies of Yukawa
structure within intersecting D-branes models. Our conclusions are
presented in section 10.

\section{{\large Fermion masses in the SM}}
The SM can be considered as a minimal theory of flavor. Being an
internally consistent renormalizable gauge theory, it has been
extremely successful in describing various experimental data. The
physical masses of the charged leptons can be directly measured
and correspond to the poles in their propagators:
\begin{equation}
M_e = 0.511 \ \makebox{MeV} \qquad M_{\mu} = 106 \ \makebox{MeV}
\qquad M_{\tau} = 1.78 \ \makebox{GeV}
\end{equation}
However, due to confinement, the quark masses cannot be directly
measured and have to be extracted from the properties of hadrons.
Various techniques are used, such as chiral perturbation theory,
QCD sum rules and lattice gauge theory. The light $u$, $d$ and
$s$ quark masses are usually normalised to the scale $\mu=1$ GeV,
corresponding to the non-perturbative scale of dynamical chiral
symmetry breaking as follows:
\begin{eqnarray}
m_u(1 \ \makebox{GeV}) & = & 4.5 \pm 1 \ \makebox{MeV} \nonumber \\
m_d(1 \ \makebox{GeV}) & = & 8 \pm 2 \ \makebox{MeV} \nonumber \\
m_s(1 \ \makebox{GeV})& = & 150 \pm 50 \ \makebox{MeV}
\end{eqnarray}
However the renormalisation scale for the heavy quark masses is
taken to be the quark mass itself:
\begin{eqnarray}
m_c(m_c) & = & 1.25 \pm 0.15 \ \makebox{GeV} \nonumber \\
m_b(m_b) & = & 4.25 \pm 0.15 \ \makebox{GeV} \nonumber \\
m_t(m_t) & = & 166 \pm 5 \ \makebox{GeV}
\end{eqnarray}
A remarkable feature of the SM is that the fermion and the gauge
boson $W^{\pm},Z$ masses have a common origin, namely the Higgs
mechanism. Fermions get masses through the Yukawa couplings to the
Higgs doublet $\phi$: \be {\cal L}_{\rm Yuk} = \la^u_{ij}
q_iCu^c_j \tilde{\phi}\,+\, \la^d_{ij} q_iCd^c_j \phi \, +
\,\la^e_{ij}l_iCe^c_j \phi ~~~~~~~~(\tilde{\phi}=i\tau_2
\phi^{\ast}) \ee so, the fermion masses are related to the weak
scale $\langle \phi \rangle=v=174$ GeV. However, the Yukawa
constants remain arbitrary: $\hat{\la}^{u,d,e}$ are general
complex $3\times 3$ matrices. The fermion mass matrices
$\hat{m}^f=\hat{\la}^fv$ ($f=u,d,e$) can be brought to the
diagonal form by the unitary transformations: \be
V_{f}^L\hat{m}^fV^R_{f}=\hat{m}^f_{diag}
 \ee
Hence, quarks are mixed in the charged current interactions: \be
{\cal
L}_W=\frac{g}{\sqrt{2}}\overline{(u_1,u_2,u_3)}_L\gamma^{\mu}W^+_{\mu}
\left( \begin{array}{c} d_1\\d_2\\d_3 \end{array} \right)_L=
\sqrt{2}g\,\overline{(u,c,t)}\,\gamma^{\mu}(1+\gamma^5)W^+_{\mu}V_{\rm
CKM} \left( \begin{array}{c} d\\s\\b \end{array} \right) \ee
where $V_{CKM}=V^{+L}_{u}V^L_{d}$ is the Cabibbo-Kobayashi-Maskawa
(CKM) matrix which represents a measure of the difference between
the unitary transformations $V_u$ and $V_d$ acting on the
left-handed up-type and down-type quarks. It has been measured:
\begin{equation}
 |V_{CKM}| = \pmatrix
{0.9734 \pm 0.0008 & 0.2196 \pm 0.0020 & 0.0036 \pm 0.0007 \cr
0.224 \pm 0.016   & 0.96 \pm 0.013   & 0.0412 \pm 0.002 \cr 0.0077
\pm 0.0014 & 0.0397 \pm 0.0033 & 0.9992 \pm 0.0002 \cr}
\end{equation}
Due to the arbitrariness in the phases of the quark fields, the
mixing matrix $V_{CKM}$ contains only one CP violating phase,
which would be of order unity $ \sin^2\delta_{CP} \sim 1 $ if the
observed CP-violating phenomena is to be attributed largely to
the CKM mechanism \cite{frog}. From solar and atmospheric neutrino
oscillation data \cite{frogref5}, we know the neutrino mass
squared differences:
\begin{equation}
 \Delta m_{21}^2 \sim 5 \times 10^{-5} \qquad
 \Delta m_{32}^2 \sim 3 \times 10^{-3}
 \label{dm2}
\end{equation}
However we do not know the absolute neutrino masses although there
is an upper limit of $m_{\nu_i} \leq 1$ eV from tritium beta decay
and from cosmology. It is of key importance that the SM exhibits
the {\em natural} suppression of the flavor changing neutral
currents (FCNC), both in the gauge boson and Higgs
exchanges\cite{berezhref15}. However, the large CKM phase creates
the strong CP problem since the overall phase of the complex
Yukawa matrices would ultimately contribute to the $\Theta$-term
in QCD and thus induce the CP violation in strong interactions,
while data suggests the bound $\Theta < 10^{-9}$.

Extensions of the SM are believed to be necessary in order to
understand the origin of the fermion masses and CKM elements.

\section{{\large Fermion masses in MSSM}}
There is no symmetry within the SM model which protects the Higgs
particle from acquiring a mass associated with physics beyond the
SM. The Higgs boson mass-squared gets corrections depending
quadratically on the SM ultra-violet cut-off $\Lambda$ from the
one-loop diagrams. Therefore the sum of the bare mass-squared
term and the radiative corrections $\sim \Lambda^2$ of the
one-loop diagrams must give a mass in the range 114 GeV $< M_h <$
200 GeV, as indicated by precision data. This leads to a
fine-tuning problem for $\Lambda >1$ TeV (the hierarchy problem).
The most popular approach to solving the hierarchy problem is
based on SUSY. Essentially all the quadratically divergent loop
diagrams have corresponding superpartner diagrams, involving
stop, gaugino and higgsino loops. In the limit of exact SUSY, the
diagrams cancel completely. However in the MSSM, it is supposed
that SUSY is softly broken at a scale $\mu = M_{SUSY} \sim 1$
TeV. So, with typical superpartner masses of order $M_{SUSY}$,
the cancellation is incomplete and $\Lambda$ is replaced by
$M_{SUSY}$.

The MSSM is consistent with the supersymmetric grand unification
of the $SU(3) \times SU(2) \times U(1)$ running gauge coupling
constants at a scale $\mu = \Lambda_{GUT} \sim 3 \times 10^{16}$
GeV, and SUSY stabilizes the hierarchy between the weak scale
$M_h$ and the grand unified one $\Lambda_{GUT}$, in the sense
that once the ratio $\frac{M_h^2}{\Lambda_{GUT}^2}$ is set to be
of order $10^{-28}$ at tree level, then it remains so to all
orders of perturbation theory. However, SUSY does not explain why
the gauge hierarchy exists in the first place.

In the MSSM the fermion masses emerge from the superpotential
terms \be {\cal W}_{\rm Yuk}= \la^u_{ij} q_i u^c_j \phi_2\,+\,
\la^d_{ij} q_i d^c_j \phi_1 \,+ \,\la^e_{ij} l_i e^c_j \phi_1,
~~~~~~~  \ee The Yukawa matrices $\la^{u,d,e,\nu}$ remain
arbitrary in the MSSM, while the presence of two Higgses $\phi_1$
and  $\phi_2$ with vacuum expectation values (VEVs)
$v_1=v\cos\beta$ and $v_2=v\sin\beta$ ($v=174$ GeV) involves also
an additional parameter $\tan\beta=v_2/v_1$.

Due to the soft SUSY breaking terms, it is apparent that the
unitary matrices which diagonalize the squark mass matrices are
not, in general, the same as the CKM matrix which diagonalize the
quark mass matrix, and unlike the SM, in the MSSM the rotations
acting on the right-handed quarks are observable especially with
non-universal soft breaking terms. These matrices work their way
into the various squark-couplings and introduce flavor
off-diagonal interactions and CP-violations beyond the CKM. These
`new' interactions are restricted by limits on rare decays, but
they might be able to explain the possible discrepancy between
the SM and experiment. In \cite{kk}, supersymmetric contributions
to the CP asymmetry of $B \to \phi K_S$ process were studied and
limits for the mixing CP asymmetry $S_{\phi K_S}$ were obtained
using the mass insertion approximation method. It was found that
the $LR$ or $RL$ mass insertion, in the terminology of the
conventional `super'-CKM basis \cite{kh99ref25}, gives the
largest contribution to $S_{\phi K_S}$, while the $LL$ or $RR$
contribution is small. Thus the large deviation between $S_{\phi
K_S}$ and $S_{J/\psi K_S}$ observed by the B--factory experiments
(Belle and BaBar) \cite{kkref2,kkref3} can be attributed to SUSY
models with large ($\sim 10^{-3}$) $LR(RL)$ mass insertions.
Similarly, the presence of non-universal A terms will be
essential for embracing the gluonic contributions to the CP
violation parameter $\eps'/\eps$ in Kaon physics, provided the
L--R mass insertions are large \cite{kkv}, and the SUSY
contribution to $\eps'/\eps$ can be of order $\sim 10^{-3}$ while
respecting the severe bound on the electric dipole moment of the
deutron.

\section{{\large Fermion masses in SUSY GUTs}}
As we said above, in the SM $\&$ MSSM the Yukawa coupling matrices
remain arbitrary, and there is no explanation to the origin of the
strong hierarchy between their eigenvalues, nor to the allignment
of the rotations acting on the up and down quarks such that the
CKM mixing angles are small. In GUTs we have more symmetries and
the question arises as to whether we gain predictivity for the
fermion masses.

Softly broken SUSY at $m_S\sim 1$ TeV is the only plausible idea
that can support the GUT against the gauge hierarchy
problem\cite{berezhref25}. On the other hand, the MSSM without
GUT is also not in best shape: unification at the string scale
gives too small $\sin^2\theta_W(M_Z)$. In SUSY GUTs we have the
following paradigm: a basic fundamental theory below the Planck
scale $M_P$ reduces to a SUSY GUT containing a compact subgroup (
$SU(5)$ or $SO(10)$), which then at $M_X\simeq 10^{16}$ GeV
breaks down to the $SU(3)\times SU(2)\times U(1)$. That is, below
the scale $M_X$ starts {\em Great Desert} with the MSSM spectrum
(see, for example, \cite{berezh} and references therein).

\subsection{{\large\bf SUSY SU(5)}}
In minimal $SU(5)$ model, the quarks and leptons of each family
fit into the multiplets \[\bar{5}_i=(d^c_i + l_i), ~~~~~
10_i=(u^c_i + q_i + e^c_i); ~~~~~~~~~~ i=1,2,3 \] and the Higgs
sector consists of chiral superfields in adjoint (24 dimensional)
representation $\Sigma$ and fundamental ($5 + \bar 5$)
representations $H$ containing the MSSM Higgs doublets
$\phi_{1,2}$ accompanied by Higgs colour triplets $\bar T,T$. The
Higgs triplets would mediate unacceptably fast proton decay via
$d=5$ operators \cite{berezhref30} unless their masses are $\sim
M_X$, in contrast to the Higgs doublets weak scale mass (DT
hierarchy problem). SUSY SU(5) offers a {\it fine tuning}
solution to the DT problem which is stable against radiative
corrections\cite{berezhref25}. However, a natural solution can be
provided by the ``missing multiplet'' mechanism\cite{berezhref31}.

As to the superpotential terms relevant for fermion masses, they
are:  \be {\cal W}_{\rm Yuk}= \la^u_{ij} 10_i H 10_j\, + \,
\la^d_{ij} 10_i \bar{H} \bar{5}_j  \ee and we get automatically $
m_b(M_X) = m_{\tau}(M_X)$ in agreement with experiment, while the
other predictions $ m_{d}/m_{s}(M_X) = m_{e}/m_{\mu}(M_X)$ fail
phenomenologically by one order of magnitude. In addition, there
is no explanation neither for the fermion mass hierarchy, nor for
the CKM mixing pattern. So it is necessary to introduce a more
complicated group theoretical structure. In fact one can consider
higher order non-renormalizable operators cut-off by the Planck
scale $M_P$ involving the 24-plet $\Sigma$, like: \be
\frac{1}{M_P}\, 10\,\Sigma H 10 \, + \, \frac{1}{M_P}\,
10\,\Sigma \bar H \bar 5 , ~~~~~~ \frac{1}{M_P^2}\, 10\, \Sigma^2
H 10 \, + \, \frac{1}{M_P^2}\, 10 \,\Sigma^2 \bar H \bar 5 ,~~
\dots \ee, which contribute to the Yukawa couplings below the
scale $M_X$ in powers of $\frac{M_X}{M_P}$. This suggests a way
out, where the renormalizable Yukawa couplings fix only the third
family masses, thus maintaining the $m_b=m_\tau$ unification,
while masses of the lighter families emerge entirely from the
higher dimensional operators, and one can avoid the {\em wrong}
prediction $m_{d,s}=m_{e,\mu}$. This is analogous to a long ago
speculated possibility that in the SM the neutrino mass is not
zero, but of order $1/M$ where $M$ would correspond to new
physics.

\subsection{{\large \bf Mass matrix textures and their origin}}
The motivation for considering mass matrix ans\"{a}tze is to
obtain testable relationships between fermion masses and mixing
angles, which might reduce the number of parameters in the Yukawa
sector and provide a hint to the physics beyond the SM. In
particular, certain elements can be put to zero (so called ``zero
textures"), and the most famous one is the Fritzch ansatze
\cite{berezhref37}: \be
\hat{\la}_{u,d}=\,\matr{0}{A}{0}{A^*}{0}{B}{0}{B^*}{C} \ee with
the assumed hierarchy of parameters
\begin{equation}
|C| \gg |B| \gg |A|
\end{equation}
Then, if we neglect the phase factors, the total number of
parameters for each matrix $\hat{\la}_{u,d}$ is reduced to 3,
i.e. just the number of quark species. This allows to express the
quark mixing angles in terms of their mass ratios: \be s_{12}=
\left|\sqrt{\frac{\la_d}{\la_s}}-
e^{i\delta}\sqrt{\frac{\la_u}{\la_c}}\right|, ~~~~
s_{23}=\left|\sqrt{\frac{\la_s}{\la_b}}-
e^{i\kappa}\sqrt{\frac{\la_c}{\la_t}}\right|,
~~~~~s_{13}=\sqrt{\frac{\la_u}{\la_c}}\, s_{23} \ee where
$\delta$ is a CP-violating phase and $\kappa$ is some unknown
phase. In particular, when $\delta\sim 1$, we obtain
$s_{12}\approx \sqrt{m_d/m_s}$ which fits the experimental value
well. However, the $s_{23}$ relation is excluded by the data for
any value of the phase $\kappa$. Consistency with experiment can,
for example, be restored by introducing a non-zero $(\la_u)_{22}$
mass matrix element \cite{frogref25}.

The origin of the textures can arise from a spontaneously broken
horizontal symmetry between the fermion families. This flavor
symmetry provides selection rules forbidding the transitions
between the various `light' left-handed and right-handed
quark-lepton states. Then, these `light' fermions can acquire
mass through higher order operators induced by the horizontal
Higgses and suitable intermediate heavy fermions mediating the
forbidden transitions (Froggatt-Nielsen mechanism). In this way,
{\em effective} SM Yukawa couplings are generated  and can be
small even though all {\em fundamental\/} Yukawa couplings of the
underlying theory are of $\cal O$(1). Such a scheme could be
implemented into SUSY GUTs ( see \cite{berezhref49and50} in the
context of $SU(5) \times SU(3)_H$), and zero textures would
appear if appropriately superheavy states are absent (see
\cite{frogref43} in the context of $SO(10)$).

\subsection{{\large \bf SUSY SO(10)}}
SO(10) grand unified theory is an appealing candidate for the
unification of quarks and leptons (family by family) and their
interactions. This is because it is the smallest group in which
each family of fermions, with a right handed neutrino, fits into a
single irrep (the spinorial representation {\bf 16}). SO(10),
thus, offers a natural explanation of the smallness of neutrino
mass through the see-saw mechanism, and consequently it
incorporates leptogenesis. Moreover, it contains the Pati-Salam
$SU(4) \times SU(2)_L \times SU(2)_R$ subgroup which is the
prototype of quark-lepton unification. Higgs fields appearing in
the ${\bf 45}_H,\ {\bf 16}_H$ and $\overline{\bf 16}_H$ are needed
to break SO(10) to the SM.  The two SM light Higgs doublets can be
accommodated by a single ${\bf 10}_H$ of SO(10), which consists of
a $5 + \bar{5}$ of SU(5) or a $(6,1,1) + (1,2,2)$ of Pati-Salam
model. Doublet-triplet splitting of the Higgs fields can be
achieved via the Dimopoulos-Wilczek mechanism \cite{albref2} if
the $\langle {\bf 45}_H \rangle$ VEV points in the $B - L$
direction. One can introduce additional Higgs fields, for example,
${\bf 126}_H,\ \overline{\bf 126}_H$ pair which can help to
stabilize the DT hierarchy. The ${\bf 126}$ Higgs representation
was used in \cite{bmsref19} to study a minimal $SO(10)$ in the
case of non--canonical see--saw, and it was shown that large
neutrino mixing angles require $b-\tau$ unification. In fact,
there are plenty of $SO(10)$ models in the literature, and they
differ by their choice of Higgs structure or the horizontal flavor
symmetry (look at \cite{alb} for a recent review). Albright and
Barr \cite{frogref43} constructed an explicit SUSY $SO(10)$ model
based on a global $U(1) \times Z_2 \times Z_2$ flavor symmetry.
The required Higgses are: one {\bf 45}, two $\mathbf{16} \oplus
\mathbf{\overline{16}}$ pairs, six {\bf 10} and five {\bf 1}. The
matter fields comprised superheavy ones in the representations:
two $\mathbf{16} \oplus \mathbf{\overline{16}}$ pairs, two {\bf
10} and six {\bf 1}. Also there is a discrete $Z_2$ matter
symmetry to avoid too rapid proton decay. The resulting mass
matrices for the down quarks and leptons are
\begin{equation}
M_D =\pmatrix{\eta       &\delta  &\delta^{\prime}e^{i\phi}\cr
          \delta         & 0           &\sigma + \epsilon/3\cr
          \delta^{\prime}e^{i\phi}  &-\epsilon/3     & 1\cr}m_b^0
\qquad M_E =\pmatrix{\eta     & \delta
&\delta^{\prime}e^{i\phi}\cr
      \delta               & 0            & -\epsilon\cr
      \delta^{\prime}e^{i\phi} &\sigma + \epsilon &1\cr}m_b^0
\end{equation}
and the up quark and Dirac neutrino mass matrices are:
\begin{equation}
M_U =\pmatrix{\eta       & 0                 & 0\cr
               0         & 0            & \epsilon/3 \cr
               0         & -\epsilon/3        & 1\cr}m_t^0
\qquad M_N =\pmatrix{\eta       & 0                 & 0\cr
               0         & 0            & \epsilon \cr
               0         & -\epsilon        & 1\cr}m_t^0
\end{equation}
Several texture zeros appear in elements for which the flavor
symmetry forces the absence of superheavy states needed to
mediate, via Froggatt-Nielsen mechanism, the corresponding
transitions. The antisymmetric $\epsilon$ terms arise from
diagrams involving the adjoint $\langle {\bf 45}_H \rangle$ Higgs
VEV pointing in the $B - L$ direction. The lopsided nature of the
large $\sigma$ terms in $D$ and $L$ arises from the appearances
of diagrams involving the $\langle \bar{5}({\bf 16}_H) \rangle$
Higgs VEV. The nine SM charged fermion masses, the three CKM
mixing angles and CP violating phase are well-fitted with the 8
parameters in the above matrices, after renormalisation group
evolution from the GUT scale.

\section{{\large \bf Fermion masses in large extra dimensions}}
This represents a new framework for solving the hierarchy problem
which does not rely on supersymmetry \cite{ahhswref1}. The
gravitational and gauge interactions unite at the electroweak
scale, and the observed weakness of gravity at long distances is
due to the existence of large new spatial dimensions. This is
particularly plausible since it could be embedded in
superstring/M theory which requires for its consistency 10/11
spacetime dimensions. However, the top-to-down approach from
strings to SM (or MSSM) has not been particularly successful as
far as phenomenology is concerned, and it may be useful to bridge
between the low energy world of the SM and the string physics at
the highest energy in a way where the presence of the extra
dimensions is somehow clearly exploited. In this scenario, the
fundamental (quantum gravity) scale $M_f$ in $4+N$ dimensional
setting can be brought far beneath the conventional Planck scale,
perhaps as low as the electroweak scale $\sim 1$ TeV. The
`effective' 4-dimensional Planck scale becomes \be M_{Pl}^2
\simeq  M_f^{2+N}R^N \ee where $R^N=V_N$ denotes the
characteristic volume of the N-dimensional compact space. With
$M_f \sim 1$ TeV close to the electroweak scale, then $R$ chosen
to reproduce the observed $M_{Pl}$ yields $R \sim
10^{\frac{30}{N}-17}$. Since the gravitational force is currently
being tested in the sub-millimeter region, this scenario is not
rejected from $N=2$ onwards.

\subsection{{\large \bf SM confined to a brane}}
In the original scenarios, the SM fields were localized on a
three-dimensional wall, or 3-brane, while the gravity alone
propagates in the extra dimensions. Such scenarios can naturally
be accommodated in string theory where the wall on which the SM
fields live can be a D-brane \cite{ahhswref2}. Remarkably, no
known lab, astrophysical or cosmological constraint excluded this
scenario \cite{ahhswref3} (see also \cite{kaisarref39} for more
stringent lower bounds on $M_f$ from supernova and neutron
stars). A number of attempts was done in order to show how the
existence of one or more extra dimensions can be exploited to
yield possible resolutions of the fermion mass hierarchy
\cite{ahhswref13}, the apparent stability of the proton
\cite{ahhswref1,ahhswref3,ahhswref11}, and so on. As an example,
not meant to be restrictive, a realistic theory of flavor in
extra dimensions was constructed in \cite{ahhsw} with $M_f$ in
the region 5-10 TeV. It is based on $U(3)^5$ flavor symmetry in
which the three Yukawa matrices are each promoted to a single
`flavon' field which propagates in the bulk and whose symmetry is
strongly broken on its `source' brane distant from us. Thus one
can understand the small flavor parameters in our world from a
hierarchy of distances in the extra dimensions. A simple brane
configuration is suggested where our 3-brane and the three source
3-branes are located on a 4-brane, so that shining occurs in 1
dimension. This makes the calculation of the Yukawa matrix
simple, and we get
\begin{equation}
\lambda_{u,d} \sim \pmatrix {\epsilon' &\epsilon'&\epsilon'\cr
\epsilon'
  & \epsilon &\epsilon\cr
\epsilon'& \epsilon & A}_{u,d},
\end{equation}
which features both a hierarchy of eigenvalues and small mixing
angles. More generally, whenever we have source triplets with
uncorrelated VEVs we are led to a CKM matrix with small mixing
angles. This basic idea can be implemented in a wide range of
models taking into account the brane geometry and the directions
in $U(3)^5$ space shone by the source triplets, and it was shown
that in a simple grid model one could give the 9 quark masses and
mixing angles quite successfully in terms of just 5 free
parameters. Note that in this picture it is essential to have a
flavor symmetry as well as a set of `source' branes with a
variety of sets of the VEVs, plus a sector of the bulk messenger
fields charged under the flavor group. The SM are localized on the
same brane and the hierarchy of the flavor symmetry breaking is
caused by locating the `source' branes at different distances.
This is to be contrasted with the attempt suggested in \cite{ds}
where the three SM families are identical but they happen to live
in 3 different branes in the extra space. The Higgs gets a VEV in
one separate `source' brane, and it decays exponentially away
from the source. In this way, the mass of the SM fermions will be
determined by the overlap of their wave functions with the Higgs
profile, and no need to postulate flavor symmetry, nor messenger
fields are needed.

\subsection{{\large \bf Universal extra dimensions: UED}}
One can study the effects of allowing some SM fields to propagate
in the extra dimensions. This could allow gauge coupling
unification \cite{acdref1}, and provide new mechanisms for SUSY
breaking \cite{acdref2} and the generation of fermion mass
hierarchies \cite{ahhswref11}, or even the existence of a Higgs
doublet \cite{acdref4}.

UED refer to a situation where the extra dimensions are
accessible to {\it all\/} the SM fields. In this case, the upper
bound from the electroweak data on the size of the extra
dimensions is significantly larger than non-UED models. This is
due to the conservation of momentum in the universal dimensions
or, in the equivalent 4-dim theory, to the conservation of the KK
number. In particular, there are no vertices involving only one
non-zero KK mode, and consequently no tree-level contributions to
the precision electroweak observables. The contributions arise
first at one loop level. In the ACD model \cite{acd}, we have one
fifth universal extra dimension compactified over $S_1/Z_2$
orbifold. There are infinite KK modes of the SM particle with
universal masses \be m_{(n)}^2 = m_0^2 + \frac{n^2}{R^2}, \ee
where $m_0$ is the mass of the zero mode which is the ordinary SM
particle. It was shown that the bound on the compactification
scale is rather large $\frac{1}{R} \geq 300$ GeV contrasted to the
non-UED models bound ($\sim$ few $TeV$) \cite{acdref5}.

Concerning the flavor/CP violation in this model, it is given, as
in the SM, by the CKM matrix only. Therefore, one would not
expect a significant deviation from the SM results in the CP
asymmetries of the $B$ decays. In fact, within the UED scenario,
the main effect of the KK modes on these processes is the
modification of the Inami-Lim one loop functions, as was found
for other processes \cite{kmref7}. In \cite{km}, it was shown
that this modification is quite limited and can not explain the
$2.7 \sigma$ deviation from $\sin 2 \beta$ in the process $B_d \to
\phi K_S$ announced by Belle and BaBar Collaborations
\cite{kmref9,kmref10}.

\section{{\large \bf Split fermions scenario}}
Another class of extra dimensions which has drawn a lot of
attention is the so called `split fermions' scenario proposed by
Arkani-Hamed and Schmaltz (AS) \cite{kmref4}. According to this
scenario, we live in a thick four dimensional subspace (thick
brane) which is infinite in the usual four spacetime dimensions
and possesses a finite volume in the extra dimensions. SM fields
are constrained to live on this thick brane whilst gravity
propagates in the whole extra dimensional spacetime; furthermore,
the Higgs and the gauge fields are free to propagate in the
entire thick brane, fermions, on the other hand, are `stuck' at
slightly different points in the extra dimensions. In this
framework, the effective four dimensional Yukawa couplings are
suppressed by exponential factors that depend on the distance
among the different fermion fields localized in the brane. Direct
coupling between fermions are exponentially suppressed by the
small overlap of their wavefunctions which are given by narrow
Gaussians, for e.g. $\psi_i \sim e^{-(y-y_i)^2/\sigma^2}$ in the
AS model of 5 dimensions, where $y_i$ is the position of the quark
$i$ in the fifth dimension and $\sigma$ is the width of its wave
function, with $\sigma \ll R$. Thus, the hierarchy of couplings
does not come from symmetry but from geometry describing the
localization of the fields in the brane with higher dimensional
couplings assumed to be of order 1.

The quark mass matrices arise from the interaction of fermions
and the VEV of the Higgs zero mode and are given by \bea
(M_u)_{ij} = \frac{v_0 (\lambda_u)_{ij}}{\sqrt{2}}
e^{\frac{-(\Delta^u_{ij})^2}{4\sigma^2}},\nonumber\\
(M_d)_{ij} = \frac{v_0 (\lambda_d)_{ij}}{\sqrt{2}}
e^{\frac{-(\Delta^d_{ij})^2}{4 \sigma^2}}, \eea where
$\Delta_{ij} = \vert y_i - y_j \vert$ is the distance between
flavor $i$ and $j$. The parameters $(\lambda_{u,d})_{ij}$ are the
5D Yukawa couplings, which are arbitrary matrices of order unity
in general. The number of free parameters is larger than the
number of the observed fermion masses and mixings, so it is easy
to accommodate the various types of Yukawa textures with
hierarchical or non-hierarchical features. Examples of
hierarchical Yukawa couplings have been obtained in
\cite{kmref20,kmref21,kmref22}, which fit all the quark and
lepton masses and mixing angles. For instance, for \be y_{Q_L}
\sim \sigma \left(\begin{array}{c}
        0\\
        14.2349\\
        8.20333
\end{array}
\right),~~~~ y_{d_R} \sim \sigma \left(\begin{array}{c}
        19.4523\\
        5.15818\\
        10.1992
\end{array}
\right),~~~~ y_{u_R} \sim \sigma \left(\begin{array}{c}
        6.13244\\
        20.092\\
        9.64483
\end{array}
\right), \label{yi} \ee
 and with $|v\la^u_{ij}|\sim 1.5$, $|v\la^u_{ij}|\sim 0.05$, one
 gets the correct fermion masses and mixings. However, the
 non-universality of the couplings with KK-gluon makes observable
 both left- and right-handed rotations ($V_L$,$V_R$) that diagonalize the mass
 matrix. In general $V_R$ matrix has six phases, and these new phases
might play an important rule in  CP violating asymmetries in the
rare $B$ decays. Also, the non-universality in the fermion
position leads to FCNC at tree level, which makes it quite
dangerous and strong bounds on the string scale were obtained:
$\frac{1}{R} > 10^4$ GeV \cite{kmref23}. In \cite{km}, the impact
of the KK contributions to the $B_d-\bar{B}_d$ mixing and CP
asymmetry of $B_d \to \phi K_S$ were analyzed, and split fermions
models were shown to be able to accommodate the Belle results
\cite{kmref9}.

In \cite{rebeloref1}, with approximately equal Yukawa strength for
both the up and down quarks, it was shown that at least two extra
dimensions were necessary in order to obtain sufficient CP
violation while reproducing the correct mass spectrum and mixing
angles.

\section{{\large \bf Fermion masses in warped geometry}}
Randall and Sundrum \cite{RS1,RS2} have proposed a mechanism
based on a non-factorizable geometry and which accounts for the
ratio between the Planck scale and the weak scales without the
need to introduce a large hierarchy between the `fundamental'
Planck scale and the volume of the compact space, as was the case
in the original large extra dimension scenarios
\cite{ahhswref1,ahhswref2,ahhswref3}. In \cite{RS1}, they assumed
a 5-dimensional non-factorizable geometry based on a slice of
AdS$_5$ spacetime with two 3-branes residing at $S_1/Z_2$
orbifold fixed points. The bulk is only populated by gravity
which is localized on one 3-brane while the SM lies on the other
3-brane. The 4-dimensional weak/Planck scale hierarchy is
generated by an exponential function of the compactification
radius, called a `warp' factor:
\[ M_W/M_{Pl} \sim e^{-M_{Pl}R} \]

Later, this picture was extended to a situation where (some of)
the SM particles reside in the five dimensional bulk
\cite{gw99,dhr00,chnoy00}. It was realized that this situation may
lead to a new flavor, and a possible geometrical interpretation
for the hierarchy of quark and lepton masses might be generated
from the warp factors. In \cite{gp}, all the SM fields live in
the 5D bulk with warped metric \be ds^2 = e^{-2\sigma(y)}
\eta_{\mu \nu} dx^{\mu} dx^{\nu} + d y^2, \ee where
$\sigma(y)=\kappa \vert y \vert$ with $\kappa \sim M_P$ is the
curvature scale. However, the Higgs field must arise from a KK
excitation that is localized by the AdS metric on the TeV-brane
in order to obtain the observable masses of the $W$ and $Z$ gauge
bosons. The fermion field can be decomposed into modes as: \be
\Psi(x,y) = \frac{1}{\sqrt{2 \pi R}} \sum_{n=0}^{\infty}
\psi^{(n)}(x) e^{2 \sigma(y)} f_n(y). \ee where $R$ is the radius
of the compactified fifth dimension. Each fermion field has a bulk
mass term parametrized by two free parameters $c_{iL}$ and
$c_{jR}$, and the fermion zero modes develop an exponential
profile: \be f_{0L(R)}(y)=\frac{e^{-c_{L(R)} \sigma(y)}}{N_0}. \ee
where $c=m_{\psi}/\kappa$ and $m_{\psi}$ is the bulk mass term.
This gives rise to four-dimensional Yukawa couplings \be
Y^{u,d}_{ij} = \frac{\la^{(5)}_{ij}\sqrt{\kappa}v_0}{\pi \kappa R}
f^{u,d}_{0iL}(\pi R) f^{u,d}_{0jR}(\pi R). \ee where
$\lambda_{ij}^{(5)}$ are the 5-dimensional Yukawas. For suitable
values of $c_{iL}$ and $c_{jR}$ one could accommodate the correct
masses and mixings. Moreover, although having fermions localized
at different points (by assuming they have different 5D masses
$c_i$) leads normally to non-universal coupling with the KK gauge
bosons, which gives rise to FCNC processes \cite{gpref22}, that
could be safely avoided in the model for the right choice of the
$c$'s. However, no choice was possible to get masses and to
suppress proton decay operators unless the Higgs is delocalized
from the TeV-barne, and here one needs SUSY.

Ref.\cite{km} also studied the flavor implications of the same
scenario and the same conclusion is reached: Because there are
plenty of free parameters (the $\la^{(5)}_{ij}$ and the $c$'s) we
can always in this class of models, as it was the case in Split
fermions scenario, obtain any type of Yukawa texture. However, the
non-universality of the gauge KK modes couplings to the fermions
is less than the non-universality in Split fermions scenario, and
it is not possible in warped geometry to deviate the value of
$S_{\phi K_S}$ from the value of $S_{J/\psi K_S}$.

\section{{\large \bf Fermion masses in heterotic string models}}
String theory is the prime candidate for the fundamental theory
of particle physics from which the SM might be derived as a
low-energy limit, and it is thought that it must be able, in
principle, to tackle fundamental questions, such as the fermion
masses and mixing, directly. In fact, in the context of
superstrings one can calculate the Yukawa couplings in terms of
scattering between the string states and certain VEVs that
parameterize the string vacuum. Some realistic superstring
derived standard-like models were constructed (e.g.
\cite{faraggi92, faraggi}) where a possible explanation of the top
quark mass hierarchy and a successful prediction of the top quark
mass were presented. However, the heterotic $E_8 \times E_8$
string theory \cite{gross} is still the most phenomenologically
promising string scenario. It consists of a 10-dim right and left
moving bosonic string, its right moving superpartner and left
moving $E_8 \times E_8$ gauge strings whose momenta span the $E_8
\times E_8$ lattice. If we bosonize the right moving fermion
string then its momenta span the $SO(10)$ weight lattice. The
theory has led to phenomenologically successful models  since one
$E_8$ factor can contain $E_6$ which in turn contains $SO(10)$
with useful subgroups such as flipped $SU(5) \times U(1)$ or
$SO(6) \times SO(4)$ while the other $E_8$ factor can be treated
as a hidden sector gauge group. In fact the 248-dim adjoint gauge
field of $E_8$ can be decomposed under $E_6 \times SU(3)$ as
$248=(78,1)+(1,8)+(27,3)+(\overline{27}+\bar{3})$ which provides,
in the first two components, gauge fields for the $E_6 \times
SU(3)$, and matter fields in $(27,3)$ with their antimatter
particles. In turn the $27$-dim representation decomposes under
$E_6 \supset SO(10)$ as $16+10+1$ providing the 16-dim spinor
representation into which the SM falls. Different schemes for
constructing classical string vacua have arisen, and it has been
possible, by compactifying 6 dimensions out of the 10 dimensions
of the theory on tori or Calaby-Yau manifolds, to build up
 four--dimensional
strings that resemble the SM in many aspects, but perhaps, the
most complete study of the Yukawa couplings has been carried out
for orbifold compactifications
\cite{cklref7,cklref9,cklref10one,cklref10two}. Abelian $Z_N$
orbifolds have been studied in depth, and we shall restrict our
discussion to them noting, however, that other orbifolds like
$Z_N \times Z_M$ have been studied
\cite{kobproduct,kobproductref11} with interesting phenomenology.

A $Z_N$ orbifold is constructed by dividing $R^6$ by a
six--dimensional lattice $\Lambda$ modded by some $Z_N$ discrete
symmetry, called the point group $P$. The space group $S$ is
defined as $S=\Lambda\times P$. There are two types of closed
strings on the orbifold. One of them is an untwisted string (U)
which is closed on torus ``unmodded" by the point group P
elements. The other is a twisted string (T) satisfying a boundary
condition \[x_{\nu}(\sigma=2\pi)=gx_{\nu}(\sigma=0),\] where $\nu
= 1, \dots , 6$ and $g$ is an element of the space group whose
point group component is non--trivial. The zero-mode, i.e. the
centre of mass coordinate, of this twisted string satisfies the
same boundary condition, and so is required to be at a fixed
point of the corresponding space group element $g$. Physical
twisted fields are associated with conjugation classes
($\{hgh^{-1},\ h\in S \}$ is the conjugation class of $g$) of the
space group rather than with particular elements, but for prime
orbifolds ($Z_N$ where $N$ is prime) conjugation classes are in
one--to--one correspondence with the fixed points of $P$.
However, for non--prime orbifolds the situation is a bit more
subtle since two different fixed points under one point group
element may be connected by another element, then both of them
correspond to the same conjugation class. We can write the space
group element $g$ in the form $g=(\theta^k, e_\nu)$, where
$\theta$ is the generator of $Z_N$ ($\theta^N=1$),
$e_v\in\Lambda$, and we have the corresponding six-dimensional
ground state $|(\theta^k,e_\nu) \rangle$. We say here that the
string belongs to the $\theta^k$ sector. The requirement of
having $N=1$ supersymmetry in four dimensions and the absence of
tachyons restrict the number of possible point groups (look at
\cite{cklref10two} for the complete list of the point groups
corresponding to $Z_N$ orbifolds). The physical states must be
invariant under a total $Z_N$ transformation which, besides the
twist $\theta$ in the 6--dimensional space, includes a $Z_N$
gauge transformation, usually represented by shifts on
$\Lambda_{E_8 \times E_8}$ and on $\Lambda_{SO(10)}$. Accordingly
one has to construct for each $\theta^k$
 sector
linear combinations of states, associated with $\theta^k$ fixed
points, that are eigenstates of $\theta$, such as $
|(\theta^k,e_\nu) \rangle + \gamma^{-1}|\theta(\theta^k,e_\nu)
\rangle + \cdots + \gamma^{-(m-1)}|\theta^{m-1}(\theta^k,e_\nu)
\rangle  $ where $\theta^m$ denotes the smallest twist fixing
$(\theta^k,e_\nu)$ itself ($m<k$). These linear combinations have
eigenvalues $\gamma =\exp[{\rm i}2 \pi p/m]$ under the
$\theta$-twist with an integer $p=0,\dots ,m-1$. Couplings
between physical states are calculated by using vertex operators
which consist of several parts: the four-dimensional part, the
six-dimensional $\theta$-eigenstate of the $\theta^k$ sector,
oscillators on it, the bosonized SO(10) part, the gauge part, and
the ghost part. Nonvanishing couplings should be invariant under
the symmetries of each part. Thus coupling terms are allowed if
they are gauge invariant and space-group invariant. Furthermore,
a product of eigenvalues $\gamma_a$ should satisfy $\prod_a
\gamma_a=1$, and the SO(10) momentum and the ghost number must be
conserved. This leads to string election rules. In
\cite{cklref10one} three point vertices satisfying the invariance
of point group and $SO(10)$ gauge group were listed for the $Z_N$
orbifolds, and it showed that the only allowed couplings are UUU,
UTT and TTT.  For the orbifold $Z_3$, however, we have only pure
untwisted (UUU) or twisted (TTT) couplings. The untwisted
renormalizable coupling is proportional to $\epsilon^{ijk}$, then
if it corresponds to the coupling of a Weinberg-Salam Higgs to
quarks (or leptons) then one of the quarks (leptons) remains
massless at this level and the other two acquire degenerate mass.
Hence, there is no mass hierarchy for the two massive generations
nor fermion mixing, even if non-canonical kinetic terms were
present \cite{cklref8one}. For other $Z_N$ orbofolds, the
situation is very similar and one can exclude the possibility of
getting mass hierarchy from untwisted matter. For the twisted
coupling, and if we denote the space group element $g$
corresponding to the fixed point $f$ in the sector $\theta^k$ by
the form $g=(\theta^k, (1-\theta^k)(f+v))$ where $v\in\Lambda$,
then the space group invariance for a Yukawa coupling to be
allowed implies that the product of the three relevant space group
elements, say $g_1 g_2 g_3$, must contain the identity, whence two
selection rules:
\begin{equation}
k_1+k_2+k_3=0\;\;{\rm mod}\ N \label{pgroup}
\end{equation}
\begin{equation}
(1-\theta^{k_1})(f_1+v_1)+\theta^{k_1}(1-\theta^{k_2})(f_2+v_2)-
(1-\theta^{k_1+k_2})(f_3+v_3)=0,\;\;v_i\in\Lambda \label{sgroup}
\end{equation}
The first one is called the point group selection rule, which
implies that the coupling must be of the form
$\theta^{k_1}\theta^{k_2} \theta^{-(k_1+k_2)}$. The second one is
called the space group selection rule and was studied in
\cite{cklref8two,cklref10one}. The lattice $\Lambda$ can get
deformations compatible with the point group and these degrees of
freedom correspond to the untwisted moduli surviving
compactification. Twisted fields at different fixed points can
communicate with each other only via world sheet instantons. The
resulting renormalizable Yukawa couplings have been calculated in
\cite{cklref7,cklref9} and they present a very rich range, which
is extremely attractive as the geometrical origin of the observed
variety and hierarchy of fermion masses. In fact, they contain
suppression factors $e^{-\alpha d^2}$ that depend on the relative
positions $d$ of the fixed points to which the fields involved in
the coupling are attached (i.e. $f_1, f_2, f_3$), and on the size
and shape of the orbifold which are given by the VEVs of moduli
fields. Thus the strength of the Yukawa is governed by the
distance from the Higgs fixed point such that the light fermions
are located further away from the Higgs. For prime orbifolds, the
space--group selection rule is of the so--called diagonal type.
This means that given two fields associated with two fixed points
$f_1, f_2$, they can only couple to a unique third fixed point
$f_3$. Let's take $Z_3$ as an illustrative example. The action of
the $Z_3$ generator $\theta$ on the $[SU(3)]^3$ root lattice
basis is: $\theta e_i=e_{i+1}$, $\theta e_{i+1}=-(e_i+e_{i+1})$
with $i=1,3,5$ denoting the three complex planes. In each 2--dim
sublattice, the twist $\theta$ acts as a rotation by
$\frac{2\pi}{3}$ and we have three $\theta$--fixed points:
\[ f^1_i=\frac{1}{3}(2e_i+e_{i+1}), \quad
f^{-1}_i=\frac{1}{3}(e_i+2e_{i+1}), \quad f^0_i=0. \] Thus we have
27$\times$ 27 = 729 allowed renormalizable Yukawa couplings
because of the diagonal--type space group selection rule. In
particular, the components of the three fixed points in each
sublattice must be either equal or different. This shows that the
Yukawa matrices derived from string theory have a very
constrained flavor structure. In \cite{cklref10two}, it was shown
that only 14 couplings out of the 729 allowed were different.
Furthermore, the lattice deformation degrees of freedom are
represented by three radii and six angles, and in the particular
case of an orthogonal lattice there are only 8 distinct couplings
which are reduced to only 4 when the radii are degenerate. In
addition to the fact that the space and point groups selection
rules are of diagonal type in prime orbifolds, the matter
associated with a given fixed point in a $\theta^k$ sector is not
degenerate, i.e. all fields have different gauge quantum numbers.
Consequently, the mass matrices in prime orbifolds are diagonal
at the renormalizable level. However, when the gauge group is
spontaneously broken after compactification there appear new
effective trilinear couplings coming from higher order
non-renormalizable operators in which some of the fields get
non--vanishing VEVs. These non-renormalizable couplings are no
longer subjected to the trilinear selection rule. For even
orbifolds, non-diagonal mass matrices at the renormalizable level
could be obtained. The reason is twofold. First, Yukawa couplings
are not necessarily of a unique $\theta^{k_1}\theta^{k_2}
\theta^{k_3}$ type. Second, the space group selection rule for a
given $\theta^{k_1}\theta^{k_2}\theta^{k_3}$ coupling is not, in
general, of the diagonal type \cite{casas92ref6}. However, as
shown in \cite{cklref8two}, the structure of the mass matrices is
still strongly constrained by the selection rules, so that, as
for prime orbifolds, no realistic prediction for the KM
parameters can emerge at the renormalizable level and one has to
call for non-renormalizable terms. Selection rules for
non-renormalizable couplings in $Z_N$ orbifolds were studied in
\cite{kob95non}, and were shown capable to lead to realistic quark
mass matrices. Under the assumption of first generation mass
coming from non-renormalizable couplings while renormalizable
couplings being responsible for the second and third generations,
we are led to the following natural ansatz for quark and lepton
mass matrices \cite{cklref8two}:
\begin{eqnarray}
M=\left( \begin {array}{ccc}
\epsilon & a & b\\
\tilde a & A & c\\
\tilde b & \tilde c & B
\end {array} \right)
\label{CM}
\end{eqnarray}
where $\epsilon,a,\tilde{a},b,\tilde{b},c,\tilde{c}<<A<<B$ in
magnitude. It was shown \cite{cklref8one,cklref8two,cklref10two}
that for a reasonable size and shape of the compactified space
the $Z_3$, $Z_4$, $Z_6$-I, and possibly $Z_7$ orbifolds can fit
the physical quark and charged-lepton masses adequately. Under
the same assumption, it was found in ref. \cite{kobproduct} that
the $Z_2 \times Z_6^{'}$, $Z_2 \times Z_6$, $Z_3 \times Z_6$ and
$Z_6 \times Z_6$ orbifolds have the possibility to derive the
Yukawa couplings for the second and third generations as well as
the measured gauge coupling constants. References
\cite{kob95,kob98} studied left-right symmetric quark mass
matrices whose up- and down-sectors have the same structure, and
derived in a $Z_6$-II orbifold model with non-renormalizable
couplings the texture: $$M_{u,d}=d_{u,d} \pmatrix{ 0 &
\varepsilon_{u,d}^3 & 0 \cr \varepsilon_{u,d}^3 &
\varepsilon_{u,d}^2 & \varepsilon_{u,d}^2 \cr 0 &
\varepsilon_{u,d}^2 & 1 \cr } \eqno(10) $$ leading to a
geometrical hierarchy $m_3m_1 \approx m_2^2$ and which can give
realistic mixing angles. In \cite{kob98}, two of the five
Ramond-Roberts-Ross types for mass matrix, with five texture zeros
in up and down quark sectors, have been derived from $Z_6$-II
orbifold models and their phenomenological consequences on flavor
mixings and $CP$ violation were analyzed at the weak scale.

In \cite{abel}, one could avoid having to use non-renormalizable
terms by assuming two Wilson lines in $Z_3$ orbifold, which then
automatically have three families of the SM particles, including
Higgses. Having three light generations of supersymmetric Higgses
introduces more Yukawa couplings, thus more flexibility in the
computation of the mass matrices, and a completely geometrical
explanation of masses and mixing is allowed. However, in order to
get the correct masses and mixings entirely at the renormalizable
level, one has to rely on the Fayet-Iliopoulos (FI) breaking.
Despite the modifications due to three Higgs families and FI
mixing, the model retains a large degree of predictivity.

\section{{\large \bf Fermion masses in intersecting branes models}}
As we said in the introduction, the advent of D-branes, as
allowed end points for open strings and which provide new ways of
realizing non abelian gauge symmetry, has increased the
phenomenological features of string theory in several respects.
Type I and Type IIB orientifold models \cite{cklref13,cklref16},
where the gauge groups of the effective low energy Lagrangian
arises from sets of coincident D branes were proposed and
investigations into their general phenomenological properties
have been possible. Yukawa textures from D-branes at
singularities were studied \cite{cklref20,cklref21}.
Nevertheless, these studies proved to be unable to explain the
experimental data, since they would generally lead to a variant
of the ``democratic" texture of Yukawa, and one has to break this
democracy by higher order or non-renormalizable operators.
However, recent studies of the flavor problem within `interscting
D-branes' models \cite{cklref24,cklref27,cklref28} seemed more
promising. In this scenario we have an interesting way to
represent the SM massless chiral fermions: a fermion sitting at
the intersection of a bunch of N D$p$-branes and another set of M
D$p$-branes ($p>3$), both containing the Minkowski space and
intersecting at some angle in the ($p-3$) extra dimensions, would
transform as $(N,\bar{M})$ under the gauge group $U(N) \times
U(M)$.

In \cite{cklref24,cklref27}, D$6$-branes were considered in type
IIA string theory compactified on a factorizable 6-torus $T^2
\times T^2 \times T^2$. If we denote the wrapping numbers of the
D6$_a$-brane on the i-th $T^2$ by $(n^i_a,m^i_a)$, then the
number of times two branes $D6_a$ and $D6_b$ intersect in $T^6$
is given by the signed intersection number
\be I_{ab}\ =\
(n_a^1m_b^1-m_a^1n_b^1)(n_a^2m_b^2-m_a^2n_b^2)(n_a^3m_b^3-m_a^3n_b^3)
\label{internumber} \ee
This gives a natural explanation of the family replication.

Two ways of embedding the SM gauge group were adopted, and both
used four stacks $a,b,c,d$ of D$6$-branes (and their orientifold
mirrors, where the mirror image of a cycle (n,m) is obtained by
reversing the second wrapping number (n,-m)). The branes are
called respectively the Baryonic, Left, Right and Leptonic branes.

In model A \cite{cklref24}, the initial gauge group is \bea Model
\ A: \qquad U(3)\times U(2) \times U(1)\times U(1)\eea with number
of branes: $N_a=3,N_b=2,N_c=1,N_d=1$. In order to yield the
desired SM spectrum, one could choose the following intersection
 numbers: \bea
I_{ab}\ & = &  \ 1 \ ;\ I_{ab*}\ =\ 2  \nonumber \\
I_{ac}\ & = &  \ -3 \ ;\ I_{ac*}\ =\ -3  \nonumber \\
I_{bd}\ & = &  \ 0  \ ;\ I_{bd*}\ =\ -3  \nonumber \\
I_{cd}\ & = &  \ -3 \ ;\ I_{cd*}\ =\ 3 \label{intersec2} \eea all
other intersections vanishing. The massless fermion spectrum is
shown in Table \ref{tableA}, where the $N_R$ represents a
right-handed neutrino. In this model one adopts the choice of
splitting the left-handed quarks into one quark $(ab)$ and two
quarks $(ab^*)$ in order to satisfy the gauge anomaly cancellation
condition which requires the same number of doublets and
antidoublets.
\begin{table}[htb] \footnotesize
\renewcommand{\arraystretch}{1.25}
\begin{center}
\begin{tabular}{|c|c|c|c|c|c|c|c|}
\hline Intersection &
 Matter fields  &   &  $Q_a$  & $Q_b $ & $Q_c $ & $Q_d$  & Y \\
\hline\hline (ab) & $Q_L$ &  $(3,\bar{2})$ & 1  & -1 & 0 & 0 & 1/6 \\
\hline (ab*) & $q_L$   &  $2( 3,2)$ &  1  & 1  & 0  & 0  & 1/6 \\
\hline (ac) & $U_R$   &  $3( {\bar 3},1)$ &  -1  & 0  & 1  & 0 & -2/3 \\
\hline (ac*) & $D_R$   &  $3( {\bar 3},1)$ &  -1  & 0  & -1  & 0 & 1/3 \\
\hline (bd*) & $ L$    &  $3(1,\bar{2})$ &  0   & -1   & 0  & -1 & -1/2  \\
\hline (cd) & $E_R$   &  $3(1,1)$ &  0  & 0  & -1  & 1  & 1   \\
\hline (cd*) & $N_R$   &  $3(1,1)$ &  0  & 0  & 1  & 1  & 0 \\
\hline \end{tabular}
\end{center} \caption{ Standard model spectrum and $U(1)$ charges
in the first model (A) \label{tableA} }
\end{table}
One can consider the possibility of NS B-flux $b^i$ on a torus
$T^i$. The existence of orientifolds on tori requires the
quantization of the $B$ flux $b = 0, 1/2$. The introduction of
non-zero NS B-flux background $b = 1/2$ on a torus modifies its
complex structure and the effect is equivalent to changing a
winding number from $(n, m)$ into $(n, m+n/2)$. The requirement to
generate the SM matter field allows only the winding numbers
denoted in Table \ref{A:winding}, where
 $\beta^i = 1 - b^i$, $\epsilon = \pm 1$ and $\rho = 1, 1/3$
 whereas $n_a^2$, $n_b^1$, $n_c^1$ and $n_d^2$ take integers.
\begin{table}[htbp]
   \centering
  \begin{tabular}[]{|c||c|c|c|}
\hline
  $N_i$ & $(n_i^1,m_i^1)$ & $(n_i^2,m_i^2)$ & $(n_i^3,m_i^3)$\\
\hline\hline $N_a = 3$ & $(1/\beta^1, 0)$ & $(n_a^2,\epsilon
\beta^2)$ &
  $(1/\rho,1/2)$\\
\hline $N_b = 2$ & $(n_b^1,-\epsilon\beta^1)$ & $(1/\beta^2, 0)$ &
$(1,
  3\rho/2)$\\
\hline
  $N_c = 1$ & $(n_c^1,3\rho\epsilon\beta^1)$ & $(1/\beta^2, 0)$ & $(0,
  1)$\\
\hline
  $N_d = 1$ & $(1/\beta^1, 0)$ & $(n_d^2, -\beta^2\epsilon/\rho)$ &
  $(1, 3\rho/2)$\\
\hline
  \end{tabular}
\caption{The winding numbers yielding the SM
  matter fields in model A}  \label{A:winding}
\end{table} In order that the hypercharge $
Q_Y = \frac{1}{6}Q_a - \frac{1}{2}Q_c + \frac{1}{2}Q_d $ remains
massless the following condition should hold:
\begin{eqnarray}
  \label{eq:condition1}
n_c^1 &=& \frac{\beta^2}{2\beta^1}(n_a^2 + 3\rho n_d^2)
\end{eqnarray} Also, one can add some D-brane stacks which do not intersect
with the first four stacks in order to satisfy the tadpole
cancellation condition:
\begin{eqnarray}
  \label{eq:condition2}
  \frac{3n_a^2}{\rho\beta^1} +   \frac{2n_b^1}{\beta^2} +
  \frac{n_d^2}{\beta^1} = 16
\end{eqnarray}
For simplicity, we can take the specific D-brane configuration
\begin{eqnarray}
\beta^1 = \beta^2 = 1,\qquad\quad\;\\
\epsilon = -1,\quad \rho = 1,\qquad\quad\\
n_a^2 = 2,\; n_b^1 = 0,\; n_c^1 = 1,\; n_d^2 = 0.
\end{eqnarray} and add
5 parallel D-branes with winding numbers $(1,0)\;(1,0)\;(1,0)$,
not intersecting with any D-brane stack, in order to satisfy the
tadpole condition (\ref{eq:condition2}) so that we end up with the
winding numbers in Table \ref{tab:concrete}.
\begin{table}
\begin{center}
  \begin{tabular}[]{|c||c|c|c|}
\hline
  $N_i$ & $(n_i^1,m_i^1)$ & $(n_i^2,m_i^2)$ & $(n_i^3,m_i^3)$\\
\hline\hline
$N_a = 3$ & $(1, 0)$ & $(2, -1)$ & $(1, 1/2)$\\
\hline
$N_b = 2$ & $(0, 1)$ & $(1, 0)$ & $(1, 3/2)$\\
\hline
$N_c = 1$ & $(1, -3)$ & $(1, 0)$ & $(0, 1)$\\
\hline
$N_d = 1$ & $(1, 0)$ & $(0, 1)$ & $(1, 3/2)$\\
\hline
  \end{tabular}
\end{center}
\caption{A combination of winding numbers yielding the SM matter
fields in model A} \label{tab:concrete}
\end{table}

In model B \cite{cklref27}, the initial gauge group is: \bea
Model \ B: \qquad U(3)\times SU(2) \times U(1)\times U(1)\eea
with number of branes: $N_a=3,N_b=1,N_c=1,N_d=1$, but $b=b^*$.
The intersection numbers are: \be\begin{array}{lcl}
I_{ab}\ = \ 3, \\
I_{ac}\ = \ -3, & & I_{ac^*}\ =\ -3, \\
I_{db}\ = \ 3, \\
I_{dc}\ = \ -3, & & I_{dc^*}\ =\ -3,
\end{array}
\label{intersecB} \ee and all others are vanishing. The massless
fermion spectrum is shown in Table \ref{mssmB}.
\begin{table}[h]
\begin{center}
\begin{tabular}{|c|c|c|c|c|c|c|}
\hline Intersection &
 SM Matter fields  & $SU(3) \times SU(2)$  &  $Q_a$   & $Q_c $ & $Q_d$  & Y \\
\hline\hline (ab) & $Q_L$ &  $3(3,2)$ & 1   & 0 & 0 & 1/6 \\
\hline (ac) & $U_R$   &  $3( {\bar 3},1)$ & -1  &  1  & 0 & -2/3 \\
\hline (ac*) & $D_R$   &  $3( {\bar 3},1)$ &  -1  & -1    & 0 &
1/3
\\
\hline (db) & $L$    &  $3(1,2)$ &  0    & 0  & 1 & -1/2  \\
\hline (dc) & $N_R$   &  $3(1,1)$ &  0  & 1  &  -1  &  0  \\
\hline (dc*) & $E_R$   &  $3(1,1)$ &  0  & -1  & -1    & 1 \\
\hline
\end{tabular}
\end{center}
 \caption{\small \label{mssmB} Standard model spectrum and $U(1)$
charges in the second model (B).}\end{table}
The model is also
anomaly free: there is no $Q_b$ anomaly condition since doublets
and antidoublets are the same in $SU(2)$. For the wrapping
numbers, one can take the values shown in Table \ref{wnumbersB1}
giving rise to the MSSM spectrum.
\begin{table}[h]
\begin{center}
\begin{tabular}{|c||c|c|c|}
\hline
 $N_i$    &  $(n_\alpha^1,m_\alpha^1)$  &  $(n_\alpha^2,m_\alpha^2)$   & $(n_\alpha^3,m_\alpha^3)$ \\
\hline\hline $N_a=3$ & $(1,0)$  &  $(1 , 3)$ &
 $(1  ,  -3)$  \\
\hline $N_b=1$ &   $(0, 1)$    &  $ (1,0)$  &
$(0,-1)$   \\
\hline $N_c=1$ & $(0,1)$  &
 $(0,-1)$  & $(1,0)$  \\
\hline $N_d=1$ &   $(1,0)$    &  $(1,3 )$  &
$(1 , -3 )$   \\
\hline
\end{tabular}
\end{center}
 \caption{\small \label{wnumbersB1} D6-brane
wrapping numbers giving rise to the chiral spectrum of the MSSM in
the second model (B). }\end{table}

As to the Higgs sector, it is complicated in model A consisting of
4 doublets $(h_i,H_i)_{i=1,2}$ at the branes intersections of
($bc,bc^*$) since we have two varieties of left quarks $(Q_L,q_L)$
and two varieties of right quarks $(U_R,D_R)$, while in model B we
have two Higgs doublets $H^u(bc)$,$H^d(bc^*)$. The Yukawa coupling
between fields $i,j,k$ is given by \be Y_{ijk} \sim e^{-A_{ijk}}
\sim \prod_{r = 1}^3 \vt \left[
\begin{array}{c}
\d^{(r)} \\ \phi^{(r)}
\end{array}
\right] (\k^{(r)}), \label{totalyuki2} \ee with $
\begin{array}{cc}
i = (i^{(1)}, i^{(2)}, i^{(3)}) ,
& \quad i^{(r)} = 0, \dots, |I_{ab}^{(r)}| -1, \\
j = (j^{(1)}, j^{(2)}, j^{(3)}) ,
& \quad j^{(r)} = 0, \dots, |I_{ca}^{(r)}| -1, \\
k = (k^{(1)}, k^{(2)}, k^{(3)})  , & \quad k^{(r)} = 0, \dots,
|I_{bc}^{(r)}| -1,
\end{array}
\label{multiindices} $ and \be \vt \left[
\begin{array}{c}
\d \\ \phi
\end{array}
\right] (\k) = \sum_{l \in \inte} e^{\pi i (\d + l)^2 \k} \
e^{2\pi i (\d + l) \phi }. \label{thetacpx} \ee with $ \d^{(r)} =
 \frac{i^{(r)}}{I_{ab}^{(r)}} + \frac{j^{(r)}}{I_{ca}^{(r)}} +
\frac{k^{(r)}}{I_{bc}^{(r)}} + \frac{d^{(r)} \cdot
\left(I_{ab}^{(r)} \eps_c^{(r)} + I_{ca}^{(r)} \eps_{b}^{(r)} +
I_{bc}^{(r)} \eps_{a}^{(r)}\right)} {I_{ab}^{(r)} I_{bc}^{(r)}
I_{ca}^{(r)}} + \frac{s^{(r)}}{d^{(r)}} \label{paramT2ncpx1}$,$
\phi^{(r)}  =  0 \label{paramT2ncpx2}$,$ \k^{(r)} =
\frac{J^{(r)}}{\a^\prime} \frac{|I_{ab}^{(r)} I_{bc}^{(r)}
I_{ca}^{(r)}|}{(d^{(r)})^2} \label{paramT2ncpx3} $ where $d^{(r)}
= g.c.d. \left( I_{ab}^{(r)}, I_{bc}^{(r)}, I_{ca}^{(r)} \right)$
and $s^{(r)} \equiv s(i^{(r)},j^{(r)},k^{(r)}) \in \inte$ is a
linear function of the integers $i^{(r)}$, $j^{(r)}$ and
$k^{(r)}$. $\eps^{(r)}$ represents the `shifts' in the r$^{th}$
torus. The exponent $A_{ijk}$ in equation (\ref{totalyuki2})
represents the area of the triangle at whose vertices lie the
fermions $i$,$j$ and $k$, and, thus, this gives a natural
exponential hierarchy for the Yukawa couplings.

One can see that Yukawa couplings in model B are always in the
form \be Y_{ij} \sim \vt^{(1)} \left[
\begin{array}{c}
\d(0) \\ 0
\end{array}
\right] \left( \k^{(1)} \right) \times \vt^{(2)} \left[
\begin{array}{c}
\d(i) \\ 0
\end{array}
\right] \left( \k^{(2)} \right) \times \vt^{(3)} \left[
\begin{array}{c}
\d(j) \\ 0
\end{array}
\right] \left( \k^{(3)} \right) \ee  so it is of a `factorizable'
form $ Y_{ij} \sim a_ib_j $ which always has two zero
eigenvalues, and the phenomenology fails because of the way in
which the family structure for the left-handed, right-handed
quarks and the Higgses is `factorized' among the different tori.
One could check that putting the quarks in one torus with one
Higgs doublet as in Table \ref{wnumbersB2} is not enough since it
lead to a Yukawa of the form $ Y_{ij}  \sim \left(
\begin{array}{ccc}
a & b & c \\
b & c & a \\
c & a & b
\end{array}
\right) $ with one degenerate mass eigenvalue.
\begin{table}[!hbp]
\begin{center}
\begin{tabular}{|c||c|c|c|}
 \hline
 $N_i$    &  $(n_\alpha^1,m_\alpha^1)$  &  $(n_\alpha^2,m_\alpha^2)$   & $(n_\alpha^3,m_\alpha^3)$ \\
\hline\hline $N_a=3$ & $(1,0)$  &  $(1 , 3)$ &
 $(1  ,  0)$  \\
\hline $N_b=1$ &   $(0, 1)$    &  $ (1,0)$  &
$(0,-1)$   \\
\hline $N_c=1$ & $(1,1)$  &
 $(1,0)$  & $(1,1)$  \\
\hline $N_d=1$ &   $(1,0)$    &  $(1,3 )$  &
$(1 , 0 )$   \\
\hline \end{tabular}
\end{center}
 \caption{\small Alternative example of D6-brane
wrapping numbers in the second model (B) leading to a chiral
spectrum of the MSSM. The family structure of both the left-handed
and right-handed quarks arises in the second torus.
\label{wnumbersB2}}\end{table}

In fact, most of intersecting brane models seem to have this acute
problem, and it seems difficult to derive non-vanishing mixing
angles as well as lighter fermion masses. However, there are
several possibilities to overcome this difficulty.

First, one can try to construct models with some further matter
content than SM fields
 by arranging the configuration of D6-branes.
Second, we can
 modify the structure of the compactified space, either by leaving the
 factorizable form $T^6 = T^2 \times T^2 \times T^2$
 and changing it into, say, $T^6 = T^4 \times T^2$, or by changing the metric on the compactified space by
 introducing a small warp factor on the tori.
The third way is to change the origin of generation of Yukawa
couplings so that it does not come from the multiple intersection
of D6-branes as in conventional models.

As an example of the first way to remedy the problem, we take the
model presented in \cite{ckl} with three supersymmetric Higgs
doublets. This model C is purely toroidal and has 4 stacks
$N_a=3$, $N_b=2$, $N_c=1$, $N_d=1$ with intersection numbers \be
\begin{array}{lccclccc}
|I_{ab}|\ = \ 3 & \mathrm{representing} & Q_L &,& |I_{bc}|=3 & \mathrm{representing} & H^u &, \\
|I_{ac}|\ = \ 3 & \mathrm{representing} & U_R &,& |I_{bd}|=6=3+3 & \mathrm{representing} & H^d,L &, \\
|I_{ad}|\ = \ 3 & \mathrm{representing} & D_R &,& |I_{cd}|=3 &
\mathrm{representing} & E_R &,
\end{array}
\label{intersecC} \ee and we can see from Table \ref{tableC},
where $\beta^2=\gamma^2=1$ and $\alpha$ is arbitrary, that the
model is anomaly free.
\begin{table}[htb] \footnotesize
\renewcommand{\arraystretch}{1.25}
\begin{center}
\begin{tabular}{|c|c|c|c|c|c|c|c|}
\hline Intersection &
 Matter fields  & $SU(3) \times SU(2)$  &  $Q_a$  & $Q_b $ & $Q_c $ & $Q_d$  & $Q_Y$ \\
\hline\hline (ab) & $Q_L$ &  $3(3,\bar{2})$ & 1  & -1 & 0 & 0 & 1/6 \\
\hline (ac) & $U_R$   &  $3( {\bar 3},1)$ &  -1  & 0  & $\beta$  & 0 & -2/3 \\
\hline (ad) & $D_R$   &  $3( {\bar 3},1)$ &  -1  & 0  & 0  & $\gamma$ & 1/3 \\
\hline (bd) & $ L$    &  $3(1,2)$ &  0   & +1   & 0  & $-\gamma$ & -1/2  \\
\hline (bd) & $ H^d_1$    &  $3(1,2)$ &  0   & +1   & 0  & $-\gamma$ & -1/2  \\
\hline (bc) & $ H^u_2$    &  $3(1,2)$ &  0   & +1   & $-\beta$  & 0 & 1/2  \\
\hline (cd) & $E_R$   &  $3(1,1)$ &  0  & 0  & $-\beta$  & $\gamma$  & 1   \\
\hline \end{tabular}
\end{center} \caption{ Standard model spectrum and $U(1)$ charges
in the third model (C) \label{tableC} }
\end{table}
The intersection numbers are obtained with the wrapping numbers
in Table (\ref{wnumbersC}), and we see that the family structure
of all the SM particles arise in the second torus.
\begin{table}[!hbp]
\begin{center}
\begin{tabular}{|c||c|c|c|}
\hline
 $N_i$    &  $(n_\alpha^1,m_\alpha^1)$  &  $(n_\alpha^2,m_\alpha^2)$   & $(n_\alpha^3,m_\alpha^3)$ \\
\hline\hline $N_a=3$ & $(1,0)$  &  $(2 , 3)$ &
 $(1  ,  0)$  \\
\hline $N_b=2$ &   $(0, 1)$    &  $ (1,3)$  &
$(0,1)$   \\
\hline $N_c=1$ & $(0,1)$  &
 $(1,0)$  & $(0,1)$  \\
\hline $N_d=1$ &   $(1,-1)$    &  $(3,3 )$  &
$(1 , -1 )$   \\
\hline \end{tabular}
\end{center}
\caption{\small Example of D6-brane wrapping numbers in the third
model (C). The family structure of the standard model particles
arises in the second torus. \label{wnumbersC}}
\end{table}
The quark Yukawa couplings for both the $H^u$ and $H^d$ Higgses
would be proportional to $ Y_{ijk} \sim \vt^{(2)} \left[
\begin{array}{c}
\d(i,j,k) \\ 0
\end{array}
\right] \left( \k^{(2)} \right) $ and we get the following
textures: \be Y_{ij1} \sim \left(
\begin{array}{ccc}
A & 0 & 0 \\
0 & 0 & B \\
0 & C & 0
\end{array}
\right), \quad Y_{ij2} \sim \left(
\begin{array}{ccc}
0 & 0 & C \\
0 & A & 0 \\
B & 0 & 0
\end{array}
\right), \quad Y_{ij3} \sim \left(
\begin{array}{ccc}
0 & B & 0 \\
C & 0 & 0 \\
0 & 0 & A
\end{array}
\right), \label{yukilocal} \ee With $\eps =\eps'$ we get the
quark mass matrices $ M^{u,d}_{ij}  =  \left(
\begin{array}{ccc}
Av^{u,d}_1 & Bv^{u,d}_3 & Cv^{u,d}_2 \\
Cv^{u,d}_3 & Av^{u,d}_2 & Bv^{u,d}_1 \\
Bv^{u,d}_2 & Cv^{u,d}_1 & Av^{u,d}_3
\end{array}
\right)  $ with $ \sum_{i=1}^3(v^u_i)^2+(v^d_i)^2 = (174)^2
(GeV)^2 \label{higgscondition} $. With the following choice of
parameters: \bea
 v^u_1  \simeq \ 63\   MeV, &v^u_2 \simeq  \ 0.95\   GeV, & v^u_3 \simeq \ 174\   GeV \nonumber \\
v^d_1  \simeq  \ 8.5\   MeV, &v^d_2 \simeq  \ 136\   MeV, & v^u_3 \simeq \ 4.2\   GeV \nonumber \\
\eps \simeq \ 0.002\ , &area \simeq  \ 18.71\  .
\label{masseparameters} \eea one gets the following quarks mass
spectrum \bea d_U&=& \{m_t\ =\ 173.9\ GeV,\ \ m_c\ =\ 1.02\ GeV,\
\ m_u\ =\ 4.3 \ MeV\}\nonumber\\d_D&=& \{m_b\ =\  4.19\ GeV,\ \
m_s\ =\ 136\ MeV,\ \ m_d\ =\
 8.2\ MeV\}\label{massresult}\eea
which are in the experimentally acceptable range, and a CKM
matrix with diagonal elements near the unity, and $(V_{CKM})_{12}
\simeq 0.216$. However, $(V_{CKM})_{13} \sim (V_{CKM})_{23} \sim
10^{-4} - 10^{-5}$. The choice $\a_{1,2} \sim v_1/v_3 \sim v_2/v_3
\sim 1$ leads to a nearly democratic Yukawa texture, while for
complex vevs and a democtratic texture, one gets the USY texture
\cite{cklref47}: $ Y^{u, d}  =  \lambda_{u,d} \left(
\begin{array}{ccc}
e^{i\varphi_{13}^{u,d}} & 1 & e^{i\varphi_{23}^{u,d}}\\
1 & e^{i\varphi_{23}^{u,d}}& e^{i\varphi_{13}^{u,d}} \\
e^{i\varphi_{23}^{u,d}}& e^{i\varphi_{13}^{u,d}} & 1
\end{array}
\right) $ which can generate the right spectrum when the phases
are small.

Now we move to the second remedy by changing the complex structure
of the internal space. In \cite{noguchi04}, the same orientifold
model A is taken but with a new feature in that a warp factor can
play an important role in generating hierarchically suppressed
fermion masses and CKM mixings. To illustrate this, let us
consider the up-quark sector and define $A_{U,ij}^{(n)}$ as the
area of a triangle formed by three intersection points $Q_{Li}$,
$U_{Rj}$ and $X_{U,i}$ on the $n$-th torus, where $Q_{Li} =
q_i\;(i = 1, 2)$, $Q_{L3} = Q_L$, $X_{U,i}= h_1\;(i = 1, 2)$ and
$X_{U,3} = H_1$. Then the Yukawa matrix for the up sector is
described as $Y_{U,ij} = \langle X_{U,i} \rangle e^{-kA_{U,ij}}$,
where $A_{U,ij} = A_{U,ij}^{(1)} + A_{U,ij}^{(3)}$. If there is no
warping, as in model A, and owing to the degeneracy of the
locations of quark doublets on the first torus, then
$A_{U,ij}^{(1)}$ depends only on the index $j$ and can be called
$A^U_j$. On the other hand, on the third torus the right handed
quarks stay at the same point, thus we define $A^Q_i =
A_{U,ij}^{(3)}$. Therefore, the Yukawa matrix for up-sector quarks
is described in a factorizable form:
\begin{eqnarray}
\label{eq:aibuj}
  Y_{U,ij} \propto e^{-kA^Q_i} e^{-kA^U_j}
\end{eqnarray}
However, if we assume in this new model D that the metric of the
first torus depends on a coordinate on the third torus $x_8$, then
the length of strings projected on the first torus becomes
dependent on the coordinate $x_8$, and so does the area swept by
the strings. In a rough approximation, we can take this effect of
the $x_8$-dependence by modifying $A^{(1)}_j$ into $ A^{U}_{j}(1 -
\epsilon_i)$, where $\epsilon_i$ is a small quantity generated by
the
 $x_8$-dependence. Surprisingly, this slight modification gives
 the realistic hierarchical structure of fermion masses and a
CKM mixing matrix. In fact, the $\epsilon_i$ should be calculable
in string theory, but if we treat it simply as an input parameter
then we reach the Yukawa matrices for the up-sector:
\begin{eqnarray}
\label{eq:yukawa}
  Y^{U}_{ij} &=& e^{-k A^{U}_{ij}},\\
  A^{U}_{ij} &=& A^Q_i + A^U_j(1 - x\, c_i).
\end{eqnarray}
In this expression, we put $\epsilon_i$ in the form $\epsilon_i =
x c_i$, where $x \ll 1$ and $c_i$ is of order 1. We define,
besides $x$, two more small quantities $y$ and $z$:
\begin{eqnarray}
  y = \frac{e^{-kA_2}}{e^{-kA_3}},\quad   z = \frac{e^{-kA_1}}{e^{-kA_2}},
\quad(A_1 > A_2 > A_3)\end{eqnarray} and we solve the eigensystem
of Yukawa matrices by series expansion with respect to $x$, $y$
and $z$ to finally obtain the leading order of the quark masses:
\begin{eqnarray}
  m_{U/D,1}^2 &=& \frac{2}{9} \beta^{U/D}_4 \left\{(c^{U/D}_3)^4 + (c^{U/D}_2)^4 y^2\right\} x^4,\\
  m_{U/D,2}^2 &=& \beta^{U/D}_0 \alpha^{U/D} (c^{U/D}_2 - c^{U/D}_3)^2 y^2 x^2,\\
  m_{U/D,3}^2 &=& \beta^{U/D}_0(1 + y^2),
\end{eqnarray}
and the CKM mixing matrix
\begin{eqnarray}
V_{\rm CKM} &\sim& \left(\matrix{1 - \epsilon^2 & \epsilon &
\epsilon^3\cr \epsilon & 1 - \epsilon^2 & \epsilon^2\cr \epsilon^3
& \epsilon^2 & 1 }\right).
\end{eqnarray}
where
\begin{eqnarray}
\alpha^U &=& \beta^U_2-(\,\beta^U_1\,)^2 , \qquad
\alpha^D \;=\;\beta^D_2 - (\,\beta^D_1\,)^2, \nonumber\\
\beta^U_0 &=& \sum_{j=1}^3 (\,b^{U}_j\,)^2, \qquad
\beta^D_0 \;=\; \sum_{j=1}^3 (\,b^{D}_j\,)^2, \nonumber\\
\beta^U_n &=& \sum_{i=1}^3 (\,b^{U}_i\,)^2 \left(\log
b^U_i\right)^n/\,\beta^U_0, \qquad
\beta^D_n \;=\; \sum_{i=1}^3 (\,b^{D}_i\,)^2 \left(\log b^D_i\right)^n/\,\beta^D_0\nonumber\\
&&\hspace{6.5cm}\mbox{(for $n = 1, 2, \cdots$)}
\end{eqnarray}
and where we assume $x$, $y$ and $z$ of order $\epsilon$. In such
a way, a hierarchical structure of a realistic CKM matrix is
obtained. It is worth noting that both large mass hierarchy and
small mixing angles can be originated from a warp factor on the
internal manifold.

\begin{figure}[th]
\includegraphics[scale=.7]{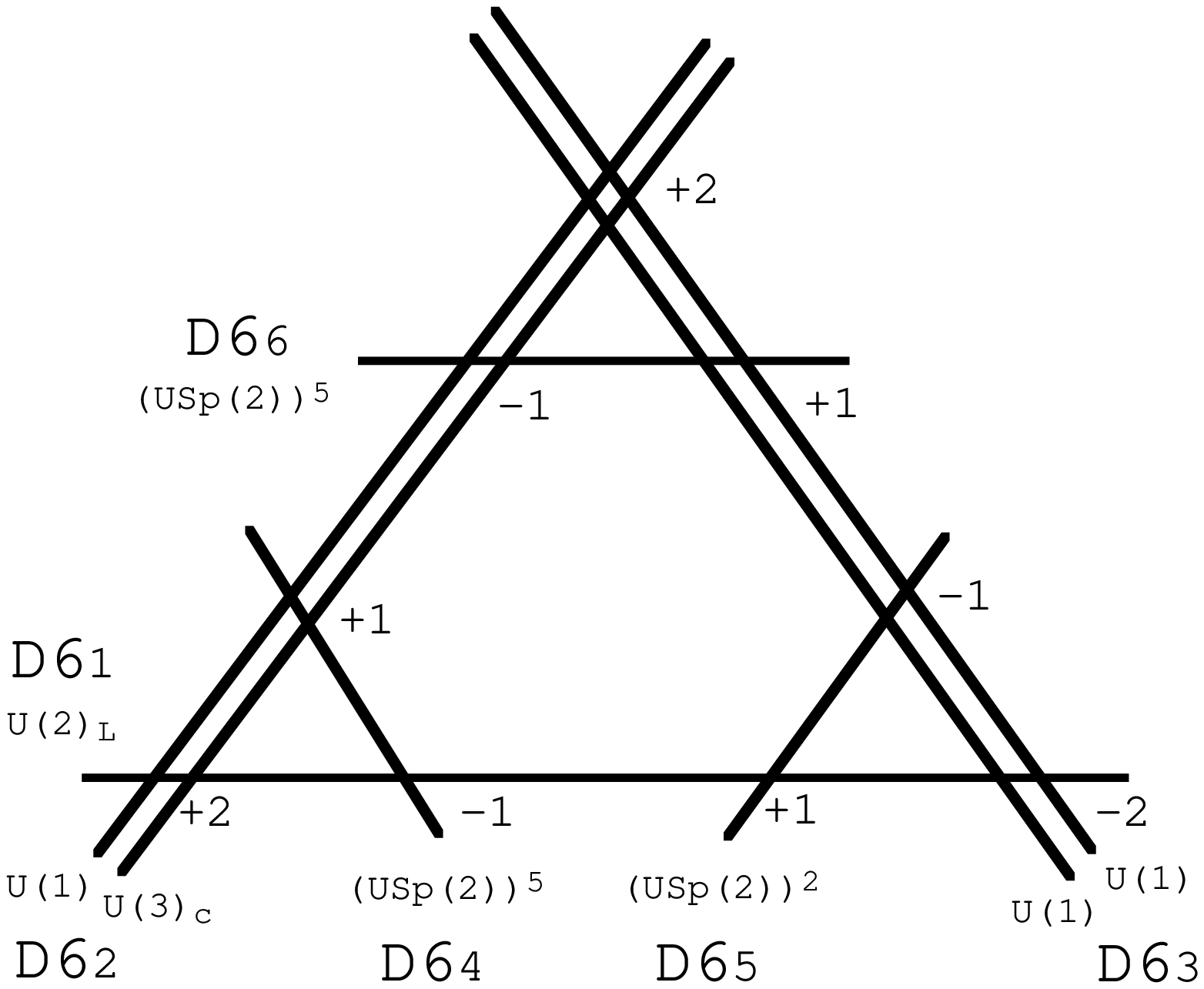}
\caption{Schematic picture of the configuration of intersecting
D6-branes in model (E). This picture describes
 only the situation of the intersection of D6-branes,
 and the relative place of each D6-brane has no meaning.
The number at the intersection point
 between D6${}_a$ and D6${}_b$ branes
 denotes intersection number $I_{ab}$ with $a<b$.}
\label{intersec}
\end{figure}

For the third way of treating the problem of factorization, we
present the model E proposed in
\cite{Kitazawa:2004nf,Kitazawa:2004ed}. It is based on
supersymmetric composite fields constructed in type IIA ${\bf
T^6}/({\bf Z_2} \times {\bf Z_2})$ orientifold
 with intersecting D6-branes. The D6-brane configuration of the model
 is given in Table \ref{config}.
\begin{table}
\begin{center}
 \begin{tabular}{|c|c|c|}
  \hline
  D6-brane & winding number & multiplicity     \\
  \hline\hline
  D6${}_1$   & $ \quad [(1,-1), (1,1), (1,0)] \quad $ & $4$ \\
  \hline
  D6${}_2$   & $ \quad [(1,1), (1,0), (1,-1)] \quad $ & $6+2$  \\
  \hline
  D6${}_3$   & $ \quad [(1,0), (1,-1), (1,1)] \quad $  & $2+2$  \\
  \hline
  D6${}_4$   & $ \quad [(1,0), (0,1), (0,-1)] \quad $ & $12$  \\
  \hline
  D6${}_5$   & $ \quad [(0,1), (1,0), (0,-1)] \quad $ & $8$ \\
  \hline
  D6${}_6$   & $ \quad [(0,1), (0,-1), (1,0)] \quad $ & $12$ \\
  \hline
 \end{tabular}
\end{center} \caption{ Configuration of intersecting D6-branes in
model (E). All three tori are considered to be rectangular
(untilted). Three D6-branes, D6${}_4$, D6${}_5$ and D6${}_6$,
 are on top of some O6-planes.
} \label{config}
\end{table}
The D6${}_2$-brane system consists of
 two parallel D6-branes with multiplicities six and two
 which are separated in the second torus
 in a consistent way with the orientifold projections. From the
 sector $aa$ of open strings ending on the same stack of branes we
 get gauge symmetries
 U$(2)_L=$SU$(2)_L \times$U$(1)_L$,
 U$(3)_c \times$U$(1) =$SU$(3)_c \times$U$(1)_c \times$U$(1)$
 and U$(1)_1 \times$U$(1)_2$ corresponding to D6${}_1$, D6${}_2$ and D6${}_3$
 branes respectively. The hypercharge is defined as
\begin{equation}
 {Y \over 2} = {1 \over 2} \left( {{Q_c} \over 3} - Q \right)
             + {1 \over 2} \left( Q_1 - Q_2 \right),
\end{equation} while the additional non-anomalous U(1) charge, $Q_R$, is defined as
\begin{equation}
 Q_R = Q_1 - Q_2.
\end{equation}
A schematic picture
 of the configuration of intersecting D6-branes of this model
 is given in figure \ref{intersec}.

There are no $ab'+b'a$, $aa'+a'a$ sectors of open string
 in this configuration. The massless particle contents are given in Table
(\ref{contents_1})
\begin{table}
\begin{center}
 \begin{tabular}{|c|c|c|}
  \hline
  sector             & $\mbox{SU}(3)_c \times \mbox{SU}(2)_L
                       \times \mbox{USp}(8)
                       \times \mbox{USp}(12)_{D6_4}
                       \times \mbox{USp}(12)_{D6_6}$
                     & field  \\
                     & ($Y/2, Q_R$)($Q_L, Q_c+Q, Q_1+Q_2$)
                     &        \\
  \hline\hline
  $D6_1 \cdot D6_2$  & $(3^*, 2, 1, 1, 1)_{(-1/6,0)(+1,-1,0)}
                         \times 2$
                     & ${\bar q}_i$ \\
                     & $(1, 2, 1, 1, 1)_{(+1/2,0)(+1,-1,0)}
                         \times 2$
                     & ${\bar l}_i$  \\
  \hline
  $D6_1 \cdot D6_4$  & $(1, 2, 1, 12, 1)_{(0,0)(-1,0,0)}$
                     & $D$ \\
  \hline
  $D6_2 \cdot D6_4$  & $(3, 1, 1, 12, 1)_{(+1/6,0)(0,+1,0)}$
                     & $C$  \\
                     & $(1, 1, 1, 12, 1)_{(-1/2,0)(0,+1,0)}$
                     & $N$  \\
  \hline
  $D6_1 \cdot D6_3$  & $(1, 2, 1, 1, 1)_{(+1/2,+1)(-1,0,+1)}
                        \times 2$
                     & $H^{(1)}_i$ \\
                     & $(1, 2, 1, 1, 1)_{(-1/2,-1)(-1,0,+1)}
                        \times 2$
                     & ${\bar H}^{(2)}_i$ \\
  \hline
  $D6_1 \cdot D6_5$  & $(1, 2, 8, 1, 1)_{(0,0)(+1,0,0)}$
                     & $T$ \\
  \hline
  $D6_3 \cdot D6_5$  & $(1, 1, 8, 1, 1)_{(+1/2,+1)(0,0,-1)}$
                     & $T^{(+)}$ \\
                     & $(1, 1, 8, 1, 1)_{(-1/2,-1)(0,0,-1)}$
                     & $T^{(-)}$ \\
  \hline
  $D6_2 \cdot D6_3$  & $(3, 1, 1, 1, 1)_{(-1/3,-1)(0,+1,-1)}
                        \times 2$
                     & ${\bar d}_i$ \\
                     & $(3, 1, 1, 1, 1)_{(+2/3,+1)(0,+1,-1)}
                        \times 2$
                     & ${\bar u}_i$ \\
                     & $(1, 1, 1, 1, 1)_{(-1,-1)(0,+1,-1)}
                        \times 2$
                     & ${\bar e}_i$ \\
                     & $(1, 1, 1, 1, 1)_{(0,+1)(0,+1,-1)}
                        \times 2$
                     & ${\bar \nu}_i$ \\
  \hline
  $D6_2 \cdot D6_6$  & $(3^*, 1, 1, 1, 12)_{(-1/6,0)(0,-1,0)}$
                     & ${\bar C}$ \\
                     & $(1, 1, 1, 1, 12)_{(+1/2,0)(0,-1,0)}$
                     & ${\bar N}$ \\
  \hline
  $D6_3 \cdot D6_6$  & $(1, 1, 1, 1, 12)_{(+1/2,+1)(0,0,+1)}$
                     & ${\bar D}^{(+)}$ \\
                     & $(1, 1, 1, 1, 12)_{(-1/2,-1)(0,0,+1)}$
                     & ${\bar D}^{(-)}$ \\
  \hline
 \end{tabular}
\end{center}
\caption{ Low-energy particle contents before ``hypercolor''
confinement in model (E). The fields from $aa$ sectors are
neglected for simplicity. } \label{contents_1}
\end{table}
where it is assumed that
 all twelve D6-branes of D6${}_4$
 are on top of one of eight O6-branes with the same winding numbers giving
 rise to USp$(12)_{{\rm D6}_4}$ gauge symmetry.
The same is also assumed
 for eight and twelve D6-branes of D6${}_5$ and D6${}_6$. The USp's gauge
 symmetries are broken
 to the factors of USp$(2)$ gauge symmetries whose interactions
 can be naturally stronger than any other unitary gauge interactions, and
 we call them ``hypercolor'' interactions.

 In the left-handed sector (D6${}_1$-D6${}_2$-D6${}_4$), the confinement of
 six USp$(2)_{{\rm D6}_4,\alpha}$ gauge interactions
 gives six generations of left-handed quark and lepton doublets:
\begin{equation}
 C_\alpha D_\alpha \sim q_\alpha,
 \qquad
 N_\alpha D_\alpha \sim l_\alpha,
\end{equation}
 where $\alpha=1,2,\cdots,6$.
Two of these six left-handed quark doublets
 and two of these six left-handed lepton doublets
 become massive through the string-level Yukawa interactions of
 the form
\begin{equation}
 W_{\rm left} =
 \sum_{i,\alpha}
  g^{\rm left-q}_{i \alpha} {\bar q}_i C_\alpha D_\alpha
 +
 \sum_{i,\alpha}
  g^{\rm left-l}_{i \alpha} {\bar l}_i N_\alpha D_\alpha,
\label{yukawa_left}
\end{equation}
 where $i=1,2$. The values of masses are given as
$g^{\rm left-q}_{i \alpha} \Lambda_L$ and $g^{\rm left-l}_{i
\alpha} \Lambda_L$, where $\Lambda_L$ denotes
 the dynamical scale of USp$(2)_{{\rm D6}_4,\alpha}$.

The same kind of confinement happens in the right-handed sector
(D6${}_2$-D6${}_3$-D6${}_6$), and the Higgs sector
(D6${}_1$-D6${}_3$-D6${}_5$), where six USp$(2)_{{\rm
D6}_6,\alpha}$ and  two USp$(2)_{{\rm D6}_5,a}$ gauge interactions
are confined to
 give six generations of right-handed quarks and leptons and eight composite
 Higgs fields, respectively:
\begin{eqnarray}
 &{\bar C}_\alpha {\bar D}^{(-)}_\alpha \sim u_\alpha,
 \qquad
 {\bar C}_\alpha {\bar D}^{(+)}_\alpha \sim d_\alpha,
 \qquad
 {\bar N}_\alpha {\bar D}^{(-)}_\alpha \sim \nu_\alpha,
 \qquad
 {\bar N}_\alpha {\bar D}^{(+)}_\alpha \sim e_\alpha&
 \\
 &T_a T^{(+)} \sim H^{(2)}_a,
 \qquad
 T_a T^{(-)} \sim {\bar H}^{(1)}_a&
\end{eqnarray}
Four generations of
 right-handed quarks and leptons and four composite Higgs fields become massive
 through the string-level Yukawa interactions of
the form:
\begin{eqnarray}
 W_{\rm right} + W_{\rm higgs}
 & = &
 \sum_{i,\alpha}
 g^{\rm right-u}_{i \alpha} {\bar u}_i {\bar C} {\bar D}^{(-)}
 +
 \sum_{i,\alpha}
 g^{\rm right-d}_{i \alpha} {\bar d}_i {\bar C} {\bar D}^{(+)}
  +
 \sum_{i,\alpha}
 g^{{\rm right-}\nu}_{i \alpha} {\bar \nu}_i {\bar N} {\bar D}^{(-)}
 \nonumber\\
 &+&
 \sum_{i,\alpha}
 g^{\rm right-e}_{i \alpha} {\bar e}_i {\bar N} {\bar D}^{(+)}
 +
 \sum_{i,a} g^{(1)}_{ia} H_i^{(1)} T_a T^{(-)}_a
 \!\!+\!\!
 \sum_{i,a} g^{(2)}_{ia} {\bar H}_i^{(2)} T_a T^{(+)}_a
\label{yukawa_right and higgs}
\end{eqnarray}
The values of masses are given as $g^{\rm right-u}_{i \alpha}
\Lambda_R$, $g^{\rm right-d}_{i \alpha} \Lambda_R$ and
$g^{(1)}_{ia} \Lambda_H$, $g^{(2)}_{ia} \Lambda_H$,
 where  $\Lambda_R$, $\Lambda_H$ are
the dynamical scales of  USp$(2)_{{\rm D6}_6,\alpha}$ and
USp$(2)_{{\rm D6}_5,a}$, respectively.

We stress here the most relevant fact about the model in that the
origin of the generation is different from the conventional
intersecting D-brane models. It
 is not the multiple intersection of D-branes,
 but the number of different D-branes with the same multiplicity and
 the same winding numbers.

The higher dimensional interactions
 in the superpotential which come from the recombination processes
 among open strings at six intersection points:
 (D6${}_2 \cdot$D6${}_4$) - (D6${}_4 \cdot$D6${}_1$)
 - (D6${}_1 \cdot$D6${}_5$) - (D6${}_5 \cdot$D6${}_3$)
 - (D6${}_3 \cdot$D6${}_6$) - (D6${}_6 \cdot$D6${}_2$), give rise to Yukawa
 interactions for the quark-lepton mass and mixing
 after the ``hypercolor'' confinement:
 \begin{eqnarray}
 \sum_{\alpha,\beta=1}^6 \sum_{a=1}^4
 {{g^u_{\alpha \beta a}} \over {M_s^3}}
  [ C_\alpha D_\alpha ]
  [ {\bar C}_\beta {\bar D}^{(-)}_\beta ]
  [ T_a T^{(+)}_a ]
 \!\!\!\!\!&\rightarrow&\!\!\!\!\!
 y^u_{\alpha \beta a}  \simeq
  g^u_{\alpha \beta a}
  {{\Lambda_L \Lambda_R \Lambda_H} \over {M_s^3}}
  \sim g^u_{\alpha \beta a}
\label{yukawa-u}
\\
 \sum_{\alpha,\beta=1}^6 \sum_{a=1}^4
 {{g^d_{\alpha \beta a}} \over {M_s^3}}
  [ C_\alpha D_\alpha ]
  [ {\bar C}_\beta {\bar D}^{(+)}_\beta ]
  [ T_a T^{(-)}_a ]
  \!\!\!\!\!&\rightarrow&\!\!\!\!\!
 y^d_{\alpha \beta a}  \simeq
  g^d_{\alpha \beta a}
  {{\Lambda_L \Lambda_R \Lambda_H} \over {M_s^3}}
  \sim g^d_{\alpha \beta a}
\label{yukawa-d}
\\
 \sum_{\alpha,\beta=1}^6 \sum_{a=1}^4
 {{g^\nu_{\alpha \beta a}} \over {M_s^3}}
  [ N_\alpha D_\alpha ]
  [ {\bar N}_\beta {\bar D}^{(-)}_\beta ]
  [ T_a T^{(+)}_a ]
 \!\!\!\!\!&\rightarrow&\!\!\!\!\!
 y^\nu_{\alpha \beta a}  \simeq
  g^\nu_{\alpha \beta a}
  {{\Lambda_L \Lambda_R \Lambda_H} \over {M_s^3}}
  \sim g^\nu_{\alpha \beta a}
\label{yukawa-nu}
\\
 \sum_{\alpha,\beta=1}^6 \sum_{a=1}^4
 {{g^e_{\alpha \beta a}} \over {M_s^3}}
  [ N_\alpha D_\alpha ]
  [ {\bar N}_\beta {\bar D}^{(+)}_\beta ]
  [ T_a T^{(-)}_a ]
 \!\!\!\!\!&\rightarrow&\!\!\!\!\!
 y^e_{\alpha \beta a}  \simeq
  g^e_{\alpha \beta a}
  {{\Lambda_L \Lambda_R \Lambda_H} \over {M_s^3}}
  \sim g^e_{\alpha \beta a}
\label{yukawa-e}
\end{eqnarray}
since all the scales of dynamics,
 $\Lambda_L$, $\Lambda_R$ and $\Lambda_H$
 are of the order of the string scale $M_s$.

 The Yukawa coupling matrices
 $g^{\rm left-q}_{i \alpha}$ in Eq.(\ref{yukawa_left}) and
 $g^{\rm right-u}_{i \alpha}$ and $g^{\rm right-d}_{i \alpha}$
 in Eq.(\ref{yukawa_right and higgs}) can be evaluated using the area law
(Eq. \ref{totalyuki2}) by:
 \begin{eqnarray}
 g^{\rm left-q} & = &
  \left(
  \begin{array}{cccccc}
  \varepsilon_3 &
  \varepsilon_2 &
  1 &
  \varepsilon_1^2 \varepsilon_2 &
  \varepsilon_1^2 &
  \varepsilon_1^2 \varepsilon_2 \varepsilon_3
  \\
  \varepsilon_1^2 \varepsilon_3 &
  \varepsilon_1^2 \varepsilon_2 &
  \varepsilon_1^2 &
  \varepsilon_2 &
  1 &
  \varepsilon_2 \varepsilon_3
  \end{array}
  \right),
\label{mass-q}\\
 g^{\rm right-u} & = &
  \left(
  \begin{array}{cccccc}
  1 &
  \varepsilon_2 &
  \varepsilon_1 &
  \varepsilon_2 \varepsilon_3^2 &
  \varepsilon_1 \varepsilon_3^2 &
  \varepsilon_1 \varepsilon_2 \varepsilon_3^2
  \\
  \varepsilon_3^2 &
  \varepsilon_2 \varepsilon_3^2 &
  \varepsilon_1 \varepsilon_3^2 &
  \varepsilon_2 & \varepsilon_1 &
  \varepsilon_1 \varepsilon_2
  \end{array}
  \right),
\label{mass-u}\\
 g^{\rm right-d} & = &
  \left(
  \begin{array}{cccccc}
  \varepsilon_1 & \varepsilon_1 \varepsilon_2 &
  1 &
  \varepsilon_1 \varepsilon_2 \varepsilon_3^2 &
  \varepsilon_3^2 &
  \varepsilon_2 \varepsilon_3^2
  \\
  \varepsilon_1 \varepsilon_3^2 &
  \varepsilon_1 \varepsilon_2 \varepsilon_3^2 &
  \varepsilon_3^2 &
  \varepsilon_1 \varepsilon_2 &
  1 &
  \varepsilon_2
  \end{array}
  \right),
\label{mass-d}
\end{eqnarray}
 where $\varepsilon_i = \exp (- A_i / 2 \pi \alpha')$ and
 $A_i$ is the $1/8$ of the area of the $i$-th torus. As for $g^u_{\alpha \beta a}$ and $g^d_{\alpha \beta a}$
 in Eqs.(\ref{yukawa-u},\ref{yukawa-d}), it is not easy to estimate them from
 first principle, however one can also apply the area law to give
 approximately an order of estimation, and obtain, say:
 \begin{eqnarray}
 g^u_{\alpha \beta a=1} & = & g^d_{\alpha \beta a=1} =
\left(
\begin{array}{cccccc}
\varepsilon_1 \varepsilon_3 & \varepsilon_1 \varepsilon_2
\varepsilon_3 & \varepsilon_1^2 \varepsilon_3 & \varepsilon_1
\varepsilon_2 \varepsilon_3 & \varepsilon_1^2 \varepsilon_3 &
\varepsilon_1^2 \varepsilon_2 \varepsilon_3 \cr \varepsilon_1
\varepsilon_2 & \varepsilon_1 \varepsilon_2^2 & \varepsilon_1^2
\varepsilon_2 & \varepsilon_1 \varepsilon_2^2 \varepsilon_3^2 &
\varepsilon_1^2 \varepsilon_2 \varepsilon_3^2 & \varepsilon_1^2
\varepsilon_2^2 \varepsilon_3^2 \cr \varepsilon_1 & \varepsilon_1
\varepsilon_2 & \varepsilon_1^2 & \varepsilon_1 \varepsilon_2
\varepsilon_3^2 & \varepsilon_1^2 \varepsilon_3^2 &
\varepsilon_1^2 \varepsilon_2 \varepsilon_3^2 \cr \varepsilon_1
\varepsilon_2 & \varepsilon_1 \varepsilon_2^2 & \varepsilon_2 &
\varepsilon_1 \varepsilon_2^2 \varepsilon_3^2 & \varepsilon_2
\varepsilon_3^2 & \varepsilon_2^2 \varepsilon_3^2 \cr
\varepsilon_1 & \varepsilon_1 \varepsilon_2 & 1 & \varepsilon_1
\varepsilon_2 \varepsilon_3^2 & \varepsilon_3^2 & \varepsilon_2
\varepsilon_3^2 \cr \varepsilon_1 \varepsilon_2 \varepsilon_3 &
\varepsilon_1 \varepsilon_2^2 \varepsilon_3 & \varepsilon_2
\varepsilon_3 & \varepsilon_1 \varepsilon_2^2 \varepsilon_3 &
\varepsilon_2 \varepsilon_3 & \varepsilon_2^2 \varepsilon_3
\end{array}
\right)_{\alpha\beta},
\end{eqnarray}
This Yukawa matrix is nontrivial in the sense that it differs from
the factorizable form, and one can check that with $\varepsilon_2
\ll \varepsilon_3 (\sim 0.01) \ll \varepsilon_1 (\sim 0.5)$ one is
led to almost realistic structure of
 quark Yukawa coupling matrices.

\section{{\large \bf Conclusions}}
The origin of the fermion masses and CKM elements is still one of
the important open questions in particle physics. Within the SM
and MSSM, we have moved a step forward in the way of
understanding the origin of mass \& relating it to the
electroweak breaking mechanism. We have seen that the fermion
mass structure in SUSY GUTs needs more analysis. Also, the other
path of giving an entirely geometrical interpretation for the
fermion masses in extra dimensions, superstrings and D-branes
scenarios is quite an interesting possibility, but we are still
far from a complete picture.


\end{document}